\documentclass[journal=jacsat,manuscript=article]{achemso}

\usepackage[version=3]{mhchem} 
\usepackage{graphicx}
\usepackage{subcaption}
\usepackage{booktabs}
\usepackage{lscape}

\author{Nia Pollard} 
\affiliation[]
{Materials Science and Engineering Division, National Institute of Standards and Technology, Gaithersburg, MD 20899, United States of America}

\email{nia.rodney-pollard@nist.gov}
\author{Kamal Choudhary}

\affiliation[]
{Materials Science and Engineering Division, National Institute of Standards and Technology, Gaithersburg, MD 20899, United States of America}

\title[]
  {BenchQC: A Benchmarking Toolkit for Quantum Computation}

\abbreviations{VQE, QC, JARVIS, CCCBDB, DFT, NISQ}

\begin{document}







\begin{abstract}
The Variational Quantum Eigensolver (VQE) is a promising algorithm for quantum computing applications in chemistry and materials science, particularly in addressing the limitations of classical methods for complex systems. This study benchmarks the performance of the VQE for calculating ground-state energies of aluminum clusters (Al\textsuperscript{-}, Al\textsubscript{2}, and Al\textsubscript{3}\textsuperscript{-}) within a quantum-density functional theory (DFT) embedding framework, systematically varying key parameters — (I) classical optimizers, (II) circuit types, (III) number of repetitions, (IV) simulator types, (V) basis sets, and (VI) noise models. Our findings demonstrate that certain optimizers achieve efficient and accurate convergence, while circuit choice and basis set selection significantly impact accuracy, with higher-level basis sets closely matching classical computation data from Numerical Python Solver (NumPy) and Computational Chemistry Comparison and Benchmark DataBase (CCCBDB). To evaluate the workflow under realistic conditions, we employed IBM noise models to simulate the effects of hardware noise. The results showed close agreement with CCCBDB benchmarks, with percent errors consistently below 0.2 \%. The results establish VQE’s capability for reliable energy estimations and highlight the importance of optimizing quantum-DFT parameters to balance computational cost and precision. This work paves the way for broader VQE benchmarking on diverse chemical systems, with plans to make results accessible on Joint Automated Repository for Various Integrated Simulations (JARVIS) and develop a Python package to support the quantum chemistry and materials science communities in advancing quantum-enhanced discovery.
\end{abstract}

\section{Introduction}
Quantum computing represents a shift in computational technology, leveraging the principles of quantum mechanics to process information in ways that classical computers cannot \cite{Nielsen_Chuang_2010, Preskill2018quantumcomputingin}. Unlike classical bits, which are binary and represent either 0 or 1, quantum bits, or qubits, can exist in a superposition of states, enabling them to perform many calculations simultaneously. This parallelism, combined with phenomena such as entanglement and quantum interference, gives quantum computers the potential to solve certain problems exponentially faster than their classical counterparts\cite{arute2019quantum, bravyi2018quantum}. Because of their potential to solve complex problems, quantum computers are at the forefront of innovation in fields like chemistry and materials science. 

The need for quantum computing in chemistry and materials discovery stems from the complexity of molecular systems and the vast number of configurations that must be explored to identify new materials\cite{aspuru2005simulated,mcardle2020quantum, bauer2020quantum}. Traditional computational methods, such as density functional theory (DFT) and post-Hartree-Fock approaches, provide valuable insights but often fall short when applied to large systems and strongly correlated electrons, or when high accuracy is required \cite{booth2013towards,motta2017towards,cohen2012challenges, alexeev2024quantum}.  Materials discovery is a field in which the identification of new compounds with desired properties, such as high-temperature superconductivity, enhanced catalytic activity, or improved energy storage, can revolutionize industries. However, the challenge lies in accurately predicting the properties of complex materials before they are synthesized \cite{rajan2015materials}. Quantum computing offers a promising avenue to overcome these challenges by enabling the precise simulation of quantum systems, allowing researchers to explore the electronic structure and properties of materials at an unprecedented level of detail. Although quantum computing has the potential to revolutionize chemistry and materials science, current noisy intermediate-scale quantum (NISQ) devices face significant limitations. These devices are constrained by noise and limited qubit counts, restricting the size of systems that can be effectively simulated using solely quantum methods\cite{Preskill2018quantumcomputingin, bharti2022noisy}. To address these challenges, quantum-DFT embedding integrates classical and quantum computing approaches, offering a potential solution that mitigates the hardware constraints of NISQ devices \cite{rossmannek2021quantum,parrish2019quantum}. 

Quantum-DFT embedding is a hybrid computational approach that combines the strengths of DFT with the precision of quantum computing\cite{bauer2020quantum,rossmannek2023quantum}. The studied system is divided into a classical region, where DFT handles the bulk of the less correlated electrons (core electrons), and a quantum region, where a quantum computer solves the more complex, strongly correlated part of the system (valence electrons). This framework allows for accurate simulations of larger and more complex systems than what current NISQ devices can handle alone. One of the key challenges in quantum chemistry is accurately capturing the behavior of strongly correlated electrons, particularly in materials with complex electronic structures. Quantum algorithms such as the Variational Quantum Eigensolver (VQE) play a crucial role in the quantum region of these simulations\cite{peruzzo2014variational,mcclean2016theory,cao2019quantum, alexeev2024quantum}. The VQE is particularly well-suited for use with NISQ devices because of its hybrid nature\cite{peruzzo2014variational}. It utilizes a classical optimizer to minimize the energy of a quantum system, represented as a parameterized quantum circuit\cite{kandala2017hardware}. By iterating between quantum measurements and classical optimization, the VQE can approximate the ground-state energy of complex systems, providing a path to more accurate and efficient simulations of molecular and material properties\cite{cao2019quantum,fedorov2022vqe}. The integration of VQE into the quantum-DFT embedding framework enables researchers to tackle challenging problems in chemistry and materials science. For example, it can help explore the electronic structure of systems with strongly correlated electrons, such as transition-metal complexes\cite{rubin2018application}. This approach offers a promising route to achieving the high precision needed for materials discovery while still mitigating the limitations of current quantum hardware. As quantum computers continue to advance, the VQE, alongside quantum-DFT embedding, could significantly enhance the predictive capabilities in chemistry and materials science, offering new insights into phenomena that were previously beyond reach.

To date, the VQE algorithm has been employed to analyze the electronic structure of smaller molecular systems such as H\textsubscript{2}, LiH, and BeH\textsubscript{2} \cite{peruzzo2014variational,mcardle2020quantum, kandala2017hardware, o2016scalable,hu2022benchmarkingvariationalquantumeigensolvers,10.1063/5.0161057}. It has also been applied to simulate electron and phonon band structures for materials applications \cite{choudhary2021quantum}.  While the VQE demonstrates significant promise for the application of quantum computing to chemical systems, there is a limited number of studies on benchmarking this method for more complex systems \cite{refId0}.  In this work, we benchmark the performance of the VQE for chemistry simulations using a previously developed quantum-DFT embedding workflow by Pollard et al\cite{pollard2025quantum}. Our results show that the VQE provides accurate results for the simulation of aluminum clusters on the quantum simulator. 
This study represents a crucial step toward expanding the applicability of the VQE beyond small molecular systems, demonstrating its potential for use in more intricate chemical simulations. Furthermore, our findings highlight the effectiveness of combining VQE with quantum-DFT embedding, a hybrid method that mitigates the limitations of current quantum devices, offering a scalable approach to the simulation of complex materials.

\section{Methodology and Computational Details}

\begin{figure}
    \centering
    \includegraphics[width=1\linewidth]{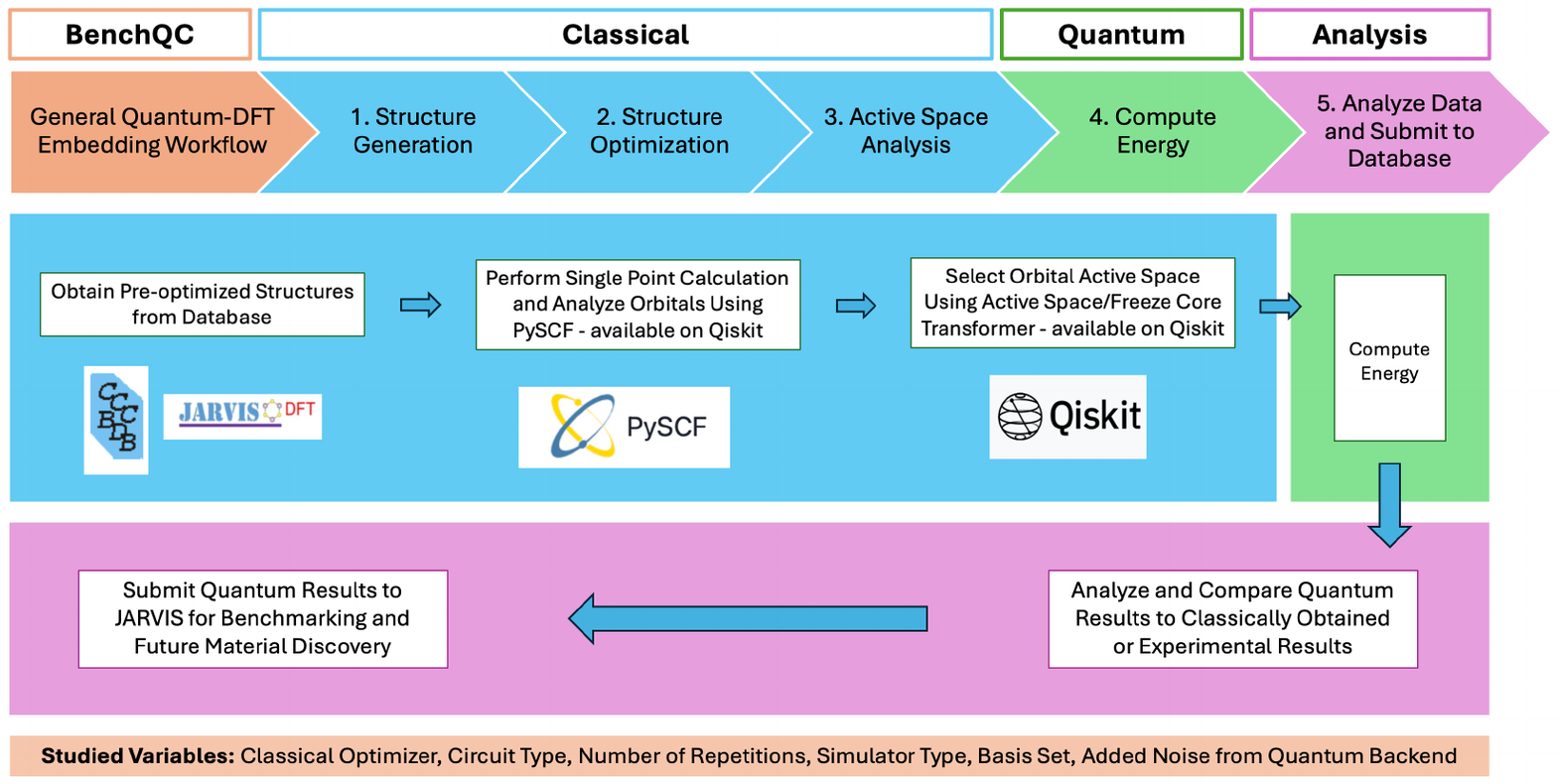}
    \caption{General Quantum-DFT Embedding Workflow: The workflow begins with structure generation where pre-optimized structures are obtained from CCCBDB, JARVIS-DFT,  or self-generated from previous studies. Single point calculations are then performed to analyze the orbital active space. Once the active space is determined, the energy can be computed using either the quantum simulator or hardware. Results are then analyzed and compared to classically obtained or experimental results.}
    \label{fig:label1}
\end{figure}
The quantum-DFT embedding workflow utilized in this study was employed on IBM's open-source platform for quantum computing, Qiskit\cite{qiskit2024} (Version 43.1). The workflow consists of five main steps as shown in Figure \ref{fig:label1}. The first step is structure generation. Pre-optimized structures are obtained from external databases, such as the Computational Chemistry Comparison and Benchmark Database (CCCBDB) \cite{iiic2022nist} and the Joint Automated Repository for Various Integrated Simulations (JARVIS-DFT)\cite{choudhary2020joint, 10.1063/5.0159299}. These databases provide the necessary starting geometries for subsequent simulations. Structures can also be obtained from previous work or they can be self-generated. After structure generation, the PySCF package \cite{10.1063/5.0006074,sun2018pyscf,sun2015libcint} (available within the Qiskit framework) is utilized to perform single-point calculations on the pre-optimized structures. This step analyzes molecular orbitals to prepare for the selection of the active space in the next phase of the workflow. The appropriate orbital active space is then determined using the Active Space Transformer available on Qiskit Nature\cite{qiskit_nature_2021}. This step is crucial for focusing the quantum computation on the most important part of the system, ensuring computational efficiency without sacrificing accuracy. The quantum region, consisting of the selected active space, is then passed to a quantum simulator or quantum hardware. The quantum computation is performed to calculate the energy of the system. Once the quantum computation is completed, the results are analyzed and compared to data obtained from Numerical Python (NumPy) or experimental results. NumPy is used as a reference for benchmarking VQE results because it performs exact diagonalization of the Hamiltonian within the defined active space and basis set. This approach yields precise ground-state energies free from noise or approximations beyond the chosen basis set, making these values reliable classical benchmarks for evaluating the accuracy of the VQE algorithm. The quantum simulation results are then submitted to the JARVIS leaderboard for benchmarking and further use in material discovery or design efforts. This integrated workflow leverages classical and quantum resources, enabling the simulation of complex materials by combining the strengths of both computational models.


In this work, we utilized both self-generated and pre-optimized aluminum molecules (ranging from Al\textsuperscript{-} to Al\textsubscript{3}\textsuperscript{-}) (Figure \ref{fig:fig3}) from a previous study.\cite{pollard2025quantum} Note that all systems with an odd number of electrons were assigned an additional negative charge. This adjustment is necessary due to a limitation of the developed workflow\cite{pollard2025quantum} and the ActiveSpaceTransformer from Qiskit Version 43.1, which requires both the active and inactive spaces to contain an even number of electrons; therefore, there cannot be any unpaired electrons. Further updates to Qiskit may enable calculations for systems with unpaired electrons. Aluminum clusters were chosen for their intermediate complexity and significant relevance to materials science, particularly in applications like catalysis, etc.\cite{mason2015aluminum, graves2005aluminum, arakawa2015reaction}. Their well-characterized properties and the availability of reliable classical benchmarks, such as those from the CCCBDB, make them ideal systems for validating quantum algorithms and exploring potential applications in materials discovery. Single-point energy calculations were performed on pre-optimized aluminum structures using the open-source software package PySCF, which is integrated as a driver within the Qiskit interface. We employed the default functional, local density approximation (LDA)\cite{kohn1965self}, as available in Qiskit Version 43.1, and varied several key parameters, including the basis set, the classical optimizer, the type of circuit, the number of repetitions (reps), and quantum simulator type. For consistency, the default parameters included a STO-3G basis set, the Sequential Least Squares Programming (SLSQP) optimizer, an EfficientSU2 ansatz with 1 repetition, and the Statevector simulator. These choices reflect commonly used settings for VQE calculations and serve as a baseline for parameter optimization. Following the initial single-point calculation, an active space of three orbitals (two filled and one unfilled) or four electrons was selected. The choice of active space is based on a previous study, which demonstrated that this configuration provided accurate results for smaller aluminum clusters\cite{pollard2025quantum}. The reduced Hamiltonian was subsequently printed, and the quantum states were encoded into qubits via Jordan Wigner Mapping. Initial calculations were performed on quantum simulators with plans to later replicate on actual quantum hardware to determine the consistency of results. 

\section{Results and discussion}
    
\subsection{I. Analyzing electronic structure with a quantum computer }
\begin{figure}
    \centering
    \includegraphics[width=0.75\linewidth]{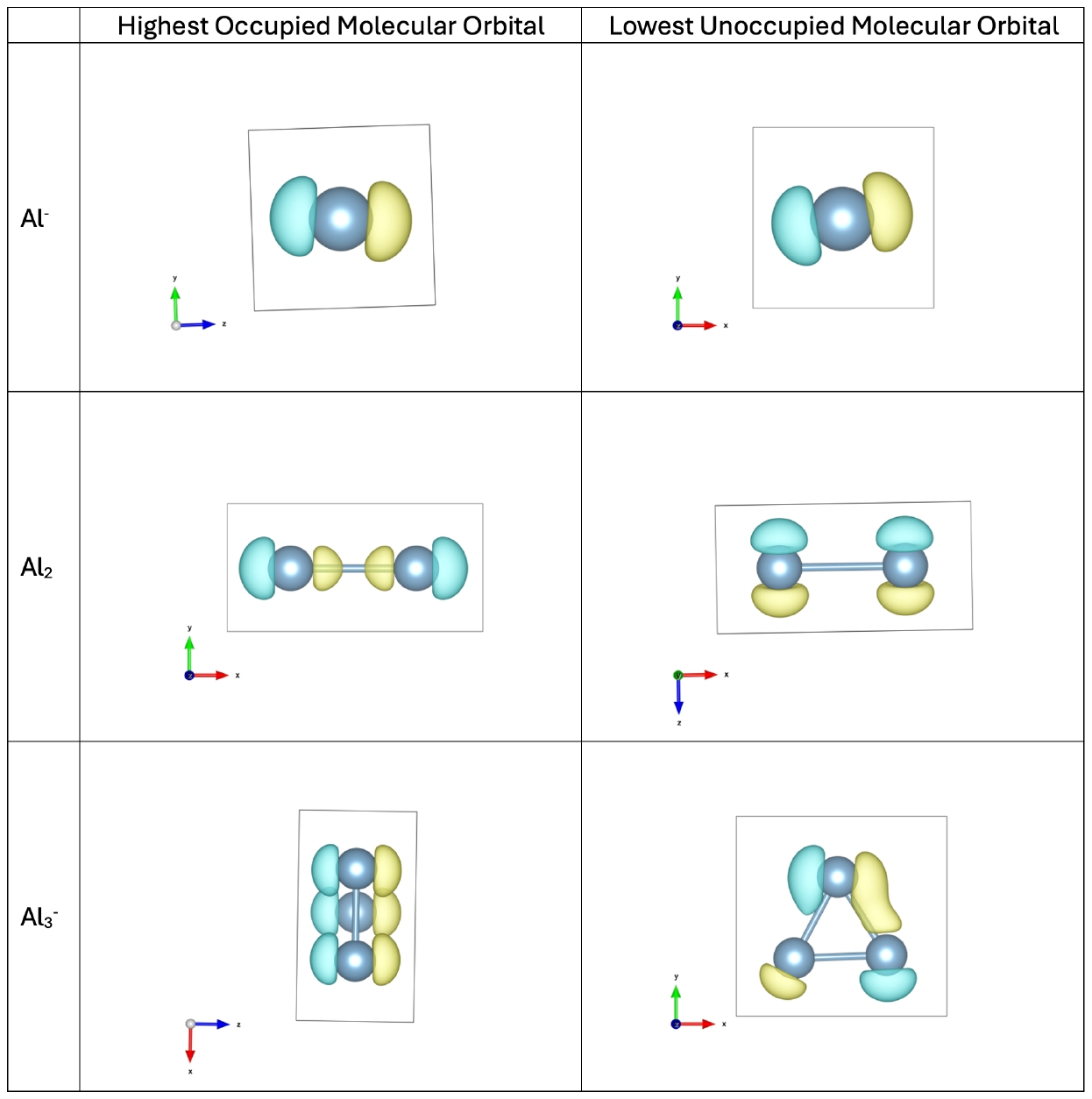}
    \caption{Visualization of  the electron density of the highest occupied molecular orbitals (HOMOs) and the lowest unoccupied molecular orbitals (LUMOs) of the studied systems with an included compass. Systems were rotated to better visualize the electron density. Results obtained using the default parameters: LDA functional, a STO-3G basis set, SLSQP optimizer, an EfficientSU2 ansatz with 1 repetition, and the Statevector simulator. }
    \label{fig:fig3}
\end{figure}

In this study, we employ quantum-DFT embedding not only to benchmark the performance of the VQE for chemical problems but also to analyze the electronic structure of the studied aluminum systems. Figure \ref{fig:fig3} presents the visualization of the electronic density of the highest occupied molecular orbitals (HOMO) and lowest unoccupied molecular orbitals (LUMO) for Al\textsuperscript{-}, Al\textsubscript{2}, and Al\textsubscript{3}\textsuperscript{-}. These visualizations were obtained using the program VESTA\cite{momma2011vesta}. Each system is rotated (as indicated by the accompanying compass) to provide a clearer view of the electron densities. The most interesting result of these visualizations is the LUMO of Al\textsubscript{3}\textsuperscript{-}, where notable electron delocalization is observed. We hypothesize that this delocalization is induced by the additional negative charge, which likely increases electron-electron repulsion within the system, thereby affecting the spatial distribution of the electrons. This behavior is consistent with the trends seen in other anionic systems, where excess negative charge can influence molecular orbitals and the distribution of electron density\cite{simons2008molecular}. 

\begin{figure}[H]
    \centering
    \includegraphics[width=1\linewidth]{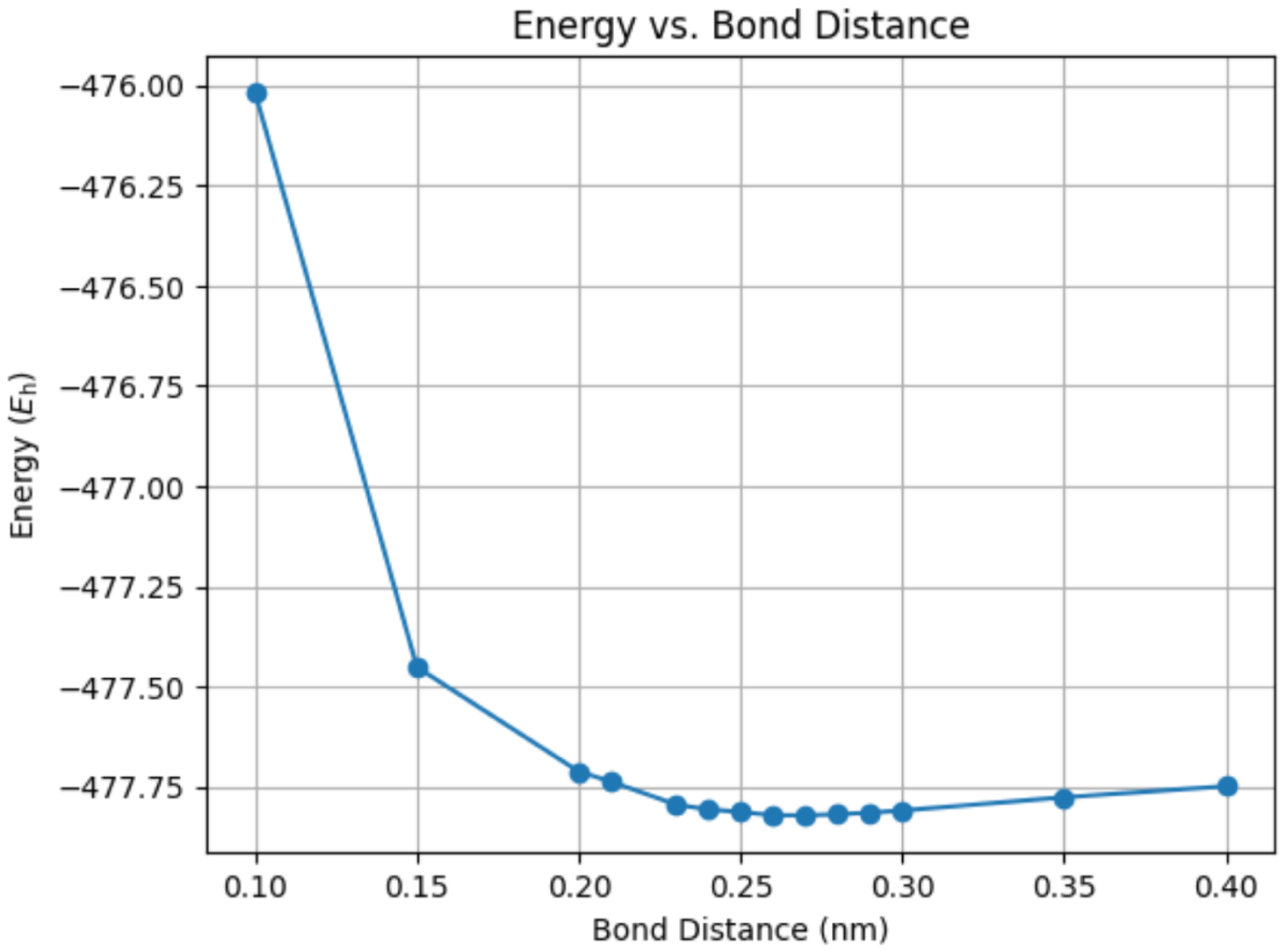}
    \caption{Ground State Energy of Al\textsubscript{2} at Various Interatomic Distances: The graph begins to converge at around 0.25 nm - 0.27 nm (2.5 \r{A} - 2.7 \r{A}), which is consistent with the typical equilibrium bond distance of Al\textsubscript{2}. VQE energy results obtained using the default parameters: LDA functional, a STO-3G basis set, SLSQP optimizer, an EfficientSU2 ansatz with 1 repetition, and the Statevector simulator. }
    \label{fig:fig4}
\end{figure}

Figure \ref{fig:fig4} shows a graph of the ground state energy vs. bond distance for Al\textsubscript{2}, where we compute the ground state energy at various interatomic distances. The shape of the plot is consistent with what is typically expected for a diatomic molecule such as Al\textsubscript{2}\cite{jensen2017introduction}. The energy drops sharply as the atoms approach each other, reaching a minimum at the equilibrium bond distance. This minimum corresponds to the most stable configuration of the molecule. Beyond this point, the energy remains relatively flat then begins to increase slightly as the bond distance continues to increase, reflecting the dissociation limit where the atoms are no longer interacting strongly. The equilibrium bond distance of Al\textsubscript{2} is typically around 0.25 nm - 0.27 nm (2.5 \r{A} - 2.7 \r{A}) \cite{bag2017stable}, which is consistent with the region where the plot reaches a minimum in our calculations.

\subsection{II. Benchmarking the performance of the VQE algorithm for chemistry problems}
To benchmark the performance of the VQE algorithm for chemistry problems, we systematically varied several key parameters, including classical optimizers, circuit type, number of reps, simulator type, and basis set. As reference values for benchmarking the VQE results, we employed the Python library NumPy for exact diagonalization of the Hamiltonian within the defined active space and basis set. We compared the results from both NumPy and our quantum calculations using different basis sets with classically obtained data from the CCCBDB.  All energies are reported in Hartree (\textit{E}\textsubscript{\textit{h}}) (see conversion to Joules used in the International System of Units\footnote{Conversion of Hartree Energy to SI Unit Joules: 1 \textit{E\textsubscript{h}} = 27.2114 eV = 4.3597e-18 J}), as it is the natural unit in quantum mechanics and widely used in quantum chemistry to ensure consistency, particularly for benchmarking studies. 

The first parameter we analyzed was the classical optimizer. In quantum-DFT embedding, classical optimizers are used to adjust parameters in hybrid quantum-classical algorithms, like VQE, to minimize the energy of the quantum region by optimizing the interaction between the quantum and classical parts of the system\cite{lavrijsen2020classical}. This ensures accurate and efficient calculations of strongly correlated regions within the framework. The optimizers we tested included Sequential Least Squares Programming (SLSQP), Limited-memory Broyden–Fletcher–Goldfarb–Shanno Bound (L\_BFGS\_B), Constrained Optimization By Linear Approximations (COBYLA), and Simultaneous Perturbation Stochastic Approximation (SPSA), with SLSQP serving as the default optimizer for all other calculations. SLSQP minimizes a scalar function subject to bounds, equality, and inequality constraints using gradient-based methods.  It converges relatively quickly with stable performance in energy minimization tasks. L\_BFGS\_B is an optimization algorithm that approximates the Broyden–Fletcher–Goldfarb–Shanno (BFGS) method, designed to handle large-scale problems with bound constraints but is often slower to converge. COBYLA is an iterative, nonlinear, derivative-free constrained optimization algorithm that uses a linear approximation approach. The algorithm is easy to use for a small number of variables. SPSA is a derivative-free optimization method that efficiently estimates gradients through random perturbations, making it well-suited for noisy functions \cite{pellow2021comparison, choudhary2021quantum}.

\begin{figure}
    \centering
    \includegraphics[width=1\linewidth]{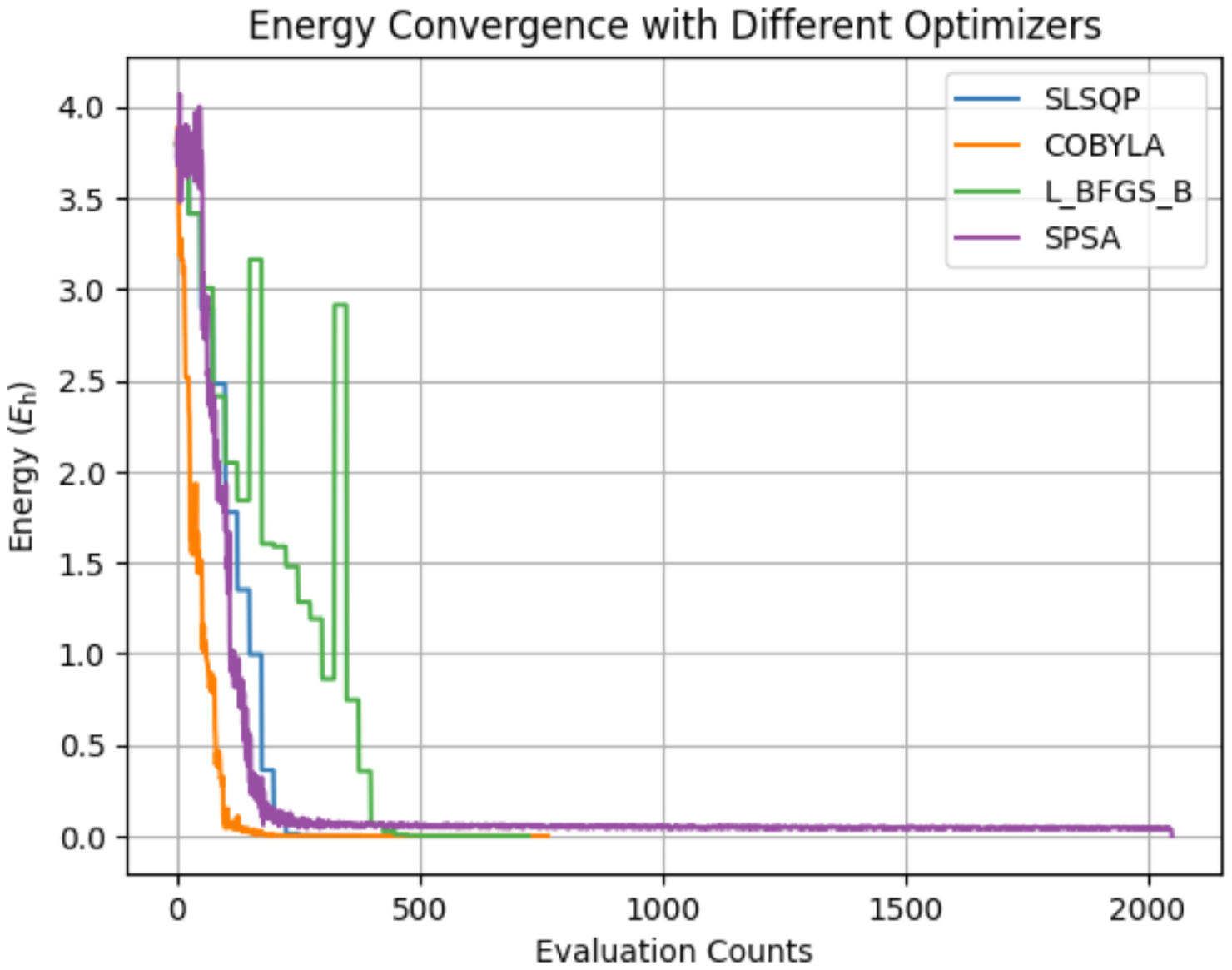}
    \caption{Energy Convergence of VQE with Varied Optimizers Rendered Using Qiskit\cite{qiskit2024}. The COBYLA optimizer appears to converge the fastest in comparison to SLSQP, L\_BFGS\_B and SPSA.}
    \label{fig:fig5}
\end{figure}

Our analysis of the different optimizers revealed that the VQE energy values differed slightly, with COBYLA and SPSA producing energy values slightly higher than those obtained with SLSQP and L\_BFGS\_B, while the NumPy values remained unchanged, as expected (Tables S1 - S3). Figure \ref{fig:fig5} presents a visual representation of the performance of each optimizer for the calculation of Al\textsubscript{2}. This analysis highlights that the COBYLA optimizer is the most efficient, converging the fastest in comparison to the other three optimizers. SLSQP performs similarly but takes a bit longer to stabilize. SPSA initially converges more slowly but stabilizes earlier than L\_BFGS\_B, which proved to be the least efficient, requiring the most steps to converge and showing notable fluctuations throughout the evaluation process. Notably, this trend was consistent across all three studied systems. These results agree with studies on other aluminum-based systems\cite{choudhary2021quantum}. 

The data table (Table \ref{tab:my_table}) further supports these observations, showing that all optimizers yield VQE energy values close to the NumPy reference values. SLSQP and L\_BFGS\_B produce the most consistent and accurate results with minimal to no deviation across all systems. This aligns with the convergence graph analysis, where SLSQP converged smoothly and L\_BFGS\_B, although slightly more erratic, still provided accurate results. The near-zero deviation also highlights the stability of these optimizers. COBYLA shows slightly more variability, with minor deviations (e.g., 0.000 037 for Al\textsuperscript{-} and 0.000 022 for Al\textsubscript{2}). Although these deviations are small and don’t significantly impact accuracy, they suggest that COBYLA may introduce minor fluctuations, which we also observed in the convergence graph where it converged quickly but with slight variations. Interestingly, despite its faster convergence in terms of energy values, COBYLA required longer job times compared to SLSQP (e.g., 16.4 seconds vs. 14.1 seconds for Al\textsuperscript{-}). This discrepancy is likely due to COBYLA’s reliance on more function evaluations per iteration. SPSA produces results that are close to the reference values but consistently show slightly higher energy values for Al\textsubscript{2} and Al\textsubscript{3}\textsuperscript{-} (Table S2 and S3).  This is consistent with the convergence graph, where SPSA initially converged more slowly and took longer to stabilize. SPSA is generally more robust in noisy conditions, but it may not be as accurate or efficient for ideal simulations. This alignment across both the convergence graph and tables indicates that SLSQP is the best choice for accuracy and stability, while COBYLA and SPSA may be preferable for applications where quick convergence or tolerance to noise is more critical.

\begin{landscape}
    
\begin{table}
\centering
\scriptsize
\caption{Mean Absolute Error of Calculated VQE Values in Comparison to NumPy Values of All Studied Systems} 
\label{tab:my_table}
\begin{tabular}{|l |l |l |l |l |l |l|} \hline   
 & $\Delta$Al\textsuperscript{-} MAE (E\textsubscript{h})& Time & $\Delta$Al\textsubscript{2} MAE (E\textsubscript{h})& Time & $\Delta$Al\textsubscript{3}\textsuperscript{-} MAE (E\textsubscript{h})& Time \\ \hline 
\textbf{Varying Classical Optimizers}&  &  &  &  &  &  \\ \hline 
\textbf{SLSQP} + Circuit 6 + 1 Rep + STO-3G + Statevector & 0.028 16 ± 0.000 000& 14.1 seconds & 0.002 92 ± 0.000 000& 11.9 seconds & 0.033 48 ± 0.000 000& 27.2 seconds \\ \hline 
\textbf{L\_BFGS\_B} + Circuit 6 + 1 Rep + STO-3G + Statevector & 0.028 18 ± 0.000 000& 14.8 seconds & 0.002 92 ± 0.000 000& 12.0 seconds & 0.029 45 ± 0.000 031& 27.7 seconds \\ \hline 
\textbf{COBLYA} + Circuit 6 + 1 Rep + STO-3G + Statevector & 0.026 80 ± 0.000 037& 16.4 seconds & 0.002 94 ± 0.000 022& 12.4 seconds & 0.018 13 ± 0.00000& 27.8 seconds \\ \hline 
\textbf{SPSA} + Circuit 6 + 1 Rep + STO-3G + Statevector & 0.027 00 ± 0.000 000& 25.1 seconds & 0.008 35 ± 0.000 000& 19.0 seconds & 0.010 32 ± 0.000 000& 35.5 seconds \\ \hline 
\textbf{Varying Circuit Type}&  &  &  &  &  &  \\ \hline 
SLSQP + \textbf{Circuit 1} + 1 Rep + STO-3G + Statevector & 0.028 02 ± 0.000 000& 11.4 seconds & 0.002 92 ± 0.000 000& 6.34 seconds & 0.018 17 ± 0.000 000& 24.0 seconds \\ \hline 
SLSQP + \textbf{Circuit 2} + 1 Rep + STO-3G + Statevector& 0.028 05 ± 0.000 000& 12.5 seconds & 0.002 92 ± 0.000 000& 8.46 seconds & 0.018 19 ± 0.000 000& 24.4 seconds \\ \hline 
SLSQP + \textbf{Circuit 3} + 1 Rep + STO-3G + Statevector & 0.028 05 ± 0.000 000& 13.1 seconds & 0.002 93 ± 0.000 000& 7.65 seconds & 0.000 000 ± 0.000 000& 24.6 seconds \\ \hline 
SLSQP + \textbf{Circuit 4} + 1 Rep + STO-3G + Statevector & 0.028 08 ± 0.000 000& 13.3 seconds & 0.002 92 ± 0.000 000& 8.68 seconds & 0.000 000 ± 0.000 000& 24.3 seconds \\ \hline 
SLSQP + \textbf{Circuit 5} + 1 Rep + STO-3G + Statevector & 0.147 52 ± 0.061 264& 14.0 seconds & 0.272 71 ± 0.097 844& 9.60 seconds & 0.029 77 ± 0.019 614& 24.0 seconds \\ \hline 
SLSQP + \textbf{Circuit 6} + 1 Rep + STO-3G + Statevector & 0.028 16 ± 0.000 000& 14.1 seconds & 0.002 92 ± 0.000 000& 11.9 seconds & 0.033 48 ± 0.000 000& 27.2 seconds \\ \hline 
\textbf{Varying Number of Repetitions}&  &  &  &  &  &  \\ \hline 
SLSQP + Circuit 6 + \textbf{1 Rep} + STO-3G + Statevector & 0.028 16 ± 0.000 000& 14.1 seconds & 0.002 92 ± 0.000 000& 11.9 seconds & 0.033 48 ± 0.000 000& 27.2 seconds \\ \hline 
SLSQP + Circuit 6 + \textbf{2 Reps} + STO-3G + Statevector & 0.027 99 ± 0.000 000& 18.8 seconds & 0.002 93 ± 0.000 000& 17.9 seconds & 0.000 01 ± 0.000 000& 34.1 seconds \\ \hline 
SLSQP + Circuit 6 + \textbf{3 Reps} + STO-3G + Statevector & 0.012 59 ± 0.000 001& 36.8 seconds & 0.002 93 ± 0.000 020& 26.9 seconds & 0.000 00 ± 0.000 000& 49.1 seconds \\ \hline 
SLSQP + Circuit 6 + \textbf{4 Reps} + STO-3G + Statevector & 0.026 76 ± 0.000 007& 64.0 seconds & 0.002 93 ± 0.000 040& 52.6 seconds & 0.000 01 ± 0.000 000& 69.0 seconds \\ \hline 
\textbf{Varying Simulator Type}&  &  &  &  &  &  \\ \hline 
SLSQP + Circuit 6 + 1 Rep + STO-3G + \textbf{Statevector }& 0.028 16 ± 0.000 000& 14.1 seconds & 0.002 92 ± 0.000 000& 11.9 seconds & 0.033 48 ± 0.000 000& 27.2 seconds \\ \hline 
SLSQP + Circuit 6 + 1 Rep + STO-3G + \textbf{QASM}& 0.559 69 ± 0.000 000& 24.5 seconds & 0.595 08 ± 0.000 000& 23.2 seconds & 0.238 41 ± 0.000 000& 42.6 seconds \\ \hline 
\textbf{Varying Basis Set}&  &  &  &  &  &  \\ \hline 
SLSQP + Circuit 6 + 1 Rep + \textbf{STO-3G} + Statevector & 0.028 16 ± 0.000 000& 14.1 seconds & 0.002 92 ± 0.000 000& 11.9 seconds & 0.033 48 ± 0.000 000& 27.2 seconds \\ \hline 
SLSQP + Circuit 6 + 1 Rep + \textbf{321-G} + Statevector & 0.007 45 ± 0.000 771& 17.2 seconds & 0.018 74 ± 0.018 338& 42.0 seconds & 0.030 05 ± 0.377 990& 71.0 seconds \\ \hline 
SLSQP + Circuit 6 + 1 Rep + \textbf{STO-6G} + Statevector & 0.026 49 ± 0.000 479& 19.8 seconds & 0.002 93 ± 0.000 000& 16.6 seconds & 0.034 26 ± 0.000 000& 52.0 seconds \\ \hline 
SLSQP + Circuit 6 + 1 Rep + \textbf{6-31G} + Statevector & 0.007 30 ± 0.000 938& 18.8 seconds & 0.003 12 ± 0.010 785& 52.2 seconds & 0.001 48 ± 0.000 000& 81.0 seconds \\ \hline 
SLSQP + Circuit 6 + 1 Rep + \textbf{cc-PVDZ} + Statevector & 0.006 78 ± 0.000 938& 22.9 seconds & 0.004 60 ± 0.016 052& 107.0 seconds &  N/A&  N/A\\ \hline 
SLSQP + Circuit 6 + 1 Rep + \textbf{cc-PVTZ} + Statevector & 0.006 35 ± 0.000 082& 60.0 seconds &  N/A&  N/A&  N/A&  N/A\\ \hline

\end{tabular}

\end{table}
\end{landscape}

\begin{figure}
    \centering
    \includegraphics[width=1\linewidth]{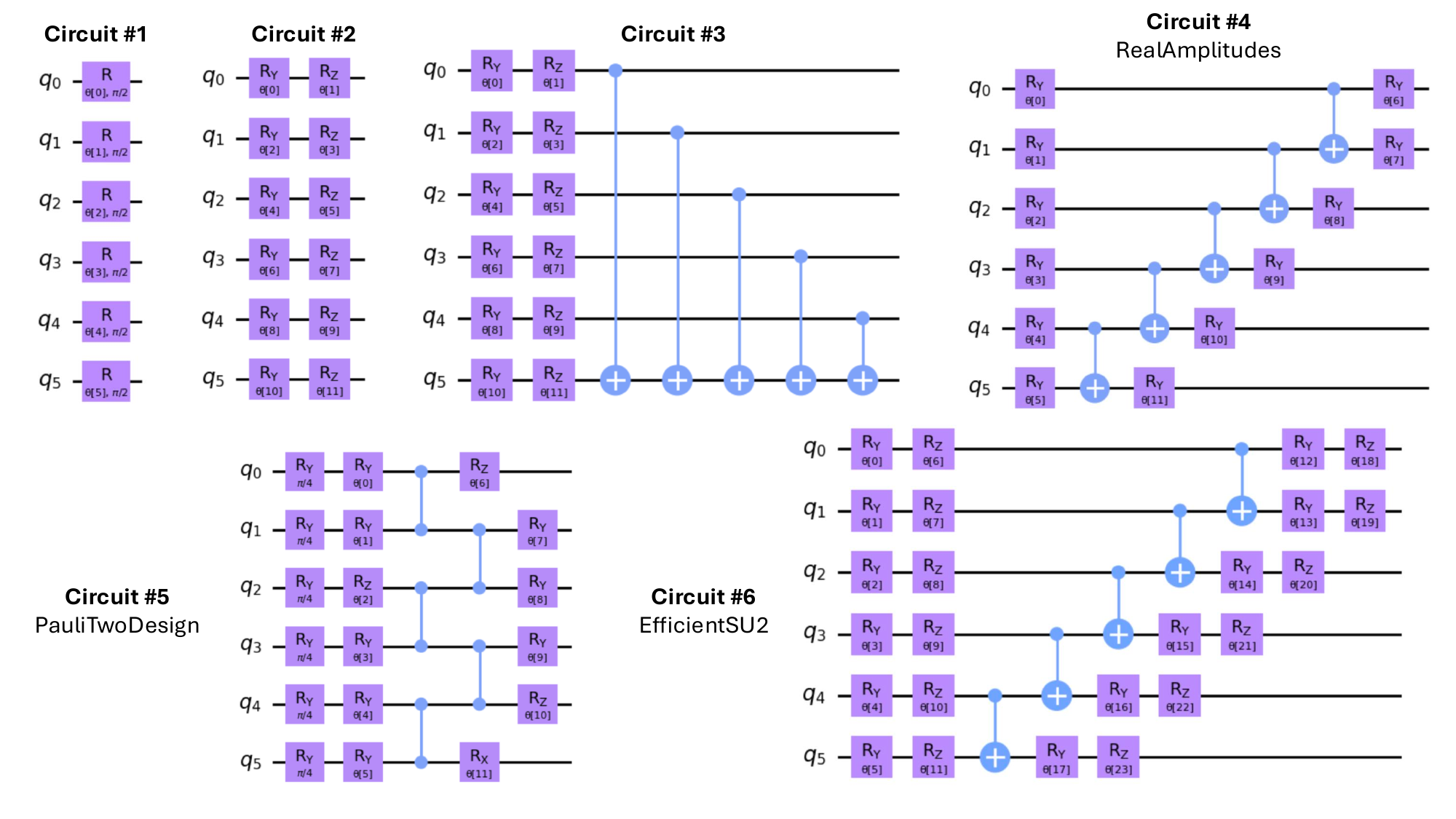}
    \caption{Visualization of Varied Circuits Rendered using Qiskit \cite{qiskit2024}: Circuits obtained from JARVIS-tools\cite{choudhary2020joint} Circuits 4, 5, and 6 are also known as RealAmplitudes, PauliTwoDesign and EfficuientSU2
circuits, respectively.}
    \label{fig:fig6}
\end{figure}

The next parameter we analyzed was the circuit type or the ansatz. The selection of quantum circuit models is largely intuitive and dependent upon the specific problem being addressed \cite{sim2019expressibility, choudhary2021quantum}. We utilized six different ansatzes for the simulation of Al\textsuperscript{-}, Al\textsubscript{2}, Al\textsubscript{3}\textsuperscript{-}, with all other parameters being held constant. An ansatz is a parameterized quantum circuit designed to approximate a quantum state \cite{mcclean2016theory, peruzzo2014variational}. The ansatz is a hypothesis about the form of the wavefunction and is optimized to minimize the energy of the system. All six circuit models discussed here are readily available in JARVIS-tools, specifically within the \textit{Jarvis.core.circuits.QuantumCircuitLibrary} module\cite{choudhary2020joint}. These circuits can all be visualized in Figure \ref{fig:fig6}. Circuits 4, 5, and 6 are also known as RealAmplitudes, PauliTwoDesign and EfficuientSU2 circuits, respectively. 

We observed that all circuits provide relatively accurate predictions of the ground state energies of the studied systems in comparison to the NumPy values (Tables S4 - S6). Circuits \#1, \#2, \#3, \#4, and \#6 show near-identical VQE values with negligible deviations across the systems, suggesting these circuits are stable and reliable for these simulations. However, Circuit \#5 (PauliTwoDesign) consistently shows a larger deviation from the NumPy reference values across all three molecules. For instance, in Al\textsuperscript{-} it has an uncertainty value of 0.061 264, in Al\textsubscript{2} an uncertainty value of 0.097 844, and in Al\textsubscript{3}\textsuperscript{-} an uncertainty value of 0.019 614 (Table \ref{tab:my_table}). This suggests that Circuit \#5 might introduce some level of approximation or noise that affects its convergence accuracy. This observation makes sense because the PauliTwoDesign ansatz is generally known for its random entangling structure and for using a fixed design based on Pauli rotations\cite{holmes2022connecting}. While this circuit can be useful for exploring highly entangled states, its structure might introduce more variability or noise in energy convergence. Circuit \#6 (EfficientSU2) on the othe   r hand shows consistent, accurate results with zero deviation across all systems, highlighting the reasons why it was the default ansatz for ground state energy calculations in the developed workflow. The slight variances between circuits indicate that the selection of the ansatz can influence the accuracy of VQE results. While most circuits perform similarly well, the differences observed, especially in Circuit \#5, emphasize the importance of carefully choosing a circuit structure that aligns with the specific properties of the system being simulated.

We also varied the number of reps, which typically refers to the number of times a quantum circuit is executed to gather enough statistics for estimating expectation values accurately. We chose to analyze only 1 to 4 reps to reduce computational cost. We found that across all three systems (Al\textsuperscript{- }, Al\textsubscript{2}, and Al\textsubscript{3}\textsuperscript{-}), the VQE energy values are very close to the reference NumPy values even with a minimal number of reps (1 rep) (Table \ref{tab:my_table}). This indicates that increasing the number of repetitions has a limited impact on accuracy, suggesting that even a simple ansatz structure can approximate the ground-state energy of these small systems well. Although additional reps (especially 3 and 4) slightly improve the energy values, the improvements are marginal. For example, in Al\textsuperscript{-} the VQE energy changes from -238.842 31 at 1 rep to -238.843 71 at 4 reps, a difference of about 0.0014 \textit{E\textsubscript{h}} (Table S7). Similarly, for Al\textsubscript{2} and Al\textsubscript{3}\textsuperscript{-} the VQE energy values remain almost unchanged across reps, suggesting diminishing returns with additional repetitions (Tables S8 - S9). The uncertainty value remains either zero or negligible for all systems and reps, with very small non-zero values appearing only at higher reps. This stability implies that the solution does not fluctuate significantly with more repetitions, showing that the ansatz structure remains consistent across reps. As expected, job times increased significantly with the number of reps. Since the energy values are so close to the NumPy values with minimal reps, using a lower number of reps (1 or 2) would be more computationally efficient without sacrificing accuracy. Higher reps (3 or 4) provide only slight improvements, so they may be unnecessary for the studied systems.

We varied the simulator type, comparing results from the Statevector simulator and the QASM (Quantum Assembly Language) simulator, both of which are readily available on the Qiskit interface. The main difference between these two simulators is that the Statevector simulator computes the exact quantum state of a system and provides deterministic results without simulating measurement, making it ideal for theoretical analysis and debugging. The QASM simulator, on the other hand, simulates measurement outcomes by running circuits multiple times (shots), providing probabilistic results like those of real quantum devices. The Statevector simulator is faster for small systems, while the QASM Simulator is more realistic for larger, noisy systems\cite{aleksandrowicz2019qiskit, mcclean2016theory}. 

In each case, the Statevector simulator produces VQE energy values closer to the reference NumPy values compared to the QASM simulator (Table S10). This indicates that the Statevector simulator provides more accurate results, likely because it does not simulate measurement noise, unlike the QASM simulator. Results from the QASM simulator deviate more significantly from the NumPy reference values. This discrepancy is expected due to the probabilistic nature of the QASM simulator, which includes measurement noise and shot-based sampling, leading to slightly higher energy values for each molecule. As the size of the system increases (from Al\textsuperscript{-} to Al\textsubscript{3}\textsuperscript{-}), the difference between QASM and Statevector results appears to increase slightly. This trend could suggest that larger systems might experience compounded effects of noise in the QASM simulator, leading to greater deviations from exact values. Simulations on the QASM simulator also required significantly longer job times compared to the Statevector simulator, reflecting the additional computational burden of simulating noise and running multiple shots (Table \ref{tab:my_table}). Despite these differences, the VQE calculations are close to the NumPy reference values in both simulators, indicating that the VQE algorithm, when optimized with the quantum-DFT embedding workflow, provides reliable energy estimates even with simulated noise. These results emphasize the trade-offs between using an ideal, noise-free simulation (Statevector) versus a more realistic but noisier simulation (QASM). It also highlights the stability of the VQE approach in estimating ground state energies, even with the limitations of current quantum devices.




We evaluated the performance of several basis sets—Slater-Type Orbital with 3 Gaussians (STO-3G), 3 Gaussian primitives with 2-1 split valence (3-21G), Slater-Type Orbital with 6 Gaussians (STO-6G), 6 Gaussian primitives with 3-1 split valence (6-31G), correlation-consistent polarized valence double-zeta (cc-pVDZ), and correlation-consistent polarized valence triple-zeta (cc-pVTZ), ordered from lowest to highest level complexity — for estimating ground-state energies of aluminum systems Al\textsuperscript{- }, Al\textsubscript{2}, and Al\textsubscript{3}\textsuperscript{-} using the VQE algorithm (Tables S11 - S13). As expected, job times increased as the basis sets became more complex (Table \ref{tab:my_table}). We note that the energy values calculated with the larger basis sets (cc-pVTZ for Al\textsubscript{2} and cc-pVDZ/cc-pVTZ for Al\textsubscript{3}\textsuperscript{-}) were not obtained due to lack of memory. We also utilized the obtained energy values of Al\textsubscript{2} for comparison against published results on the CCCBDB\cite{cccbdb}, which were calculated using DFT and LDA (Table \ref{tab:my_table16}). This comparative data is important to this study because it validates the accuracy of the developed quantum-DFT embedding workflow and proves that the VQE is a viable method for chemistry and materials science analysis. 

The results show a general trend across the basis sets: as the basis set quality increases, the VQE energy values approach the NumPy reference values, indicating improved accuracy with more complex basis sets. For instance, in the Al\textsuperscript{- } system, STO-3G provides a VQE energy of -238.842 31 \textit{E\textsubscript{h}}, while cc-pVDZ gives a closer approximation at -241.860 32 \textit{E\textsubscript{h}}, aligning well with the NumPy reference of -241.866 67 \textit{E}\textsubscript{\textit{h}} (Table S11). Similarly, for Al\textsubscript{2} and Al\textsubscript{3}\textsuperscript{-}, the higher-level basis sets (6-31G and cc-pVDZ) yield results that more closely match the reference values, underscoring the importance of basis set choice in capturing accurate ground state energies for aluminum clusters (Tables S12 - S13). When analyzing the percent errors among the basis sets in comparison to CCCBDB (Table \ref{tab:my_table16}), we see a consistent decrease in error as the basis set quality improves. For Al\textsubscript{2}, the STO-3G basis set produces a percent error of 0.1369 \%, whereas the cc-pVDZ basis set reduces this error to 0.1119 \%. This trend suggests that while minimal basis sets like STO-3G are computationally efficient, they sacrifice some accuracy. In contrast, the larger and more complete basis sets, such as cc-pVDZ, offer more reliable estimates of electronic energies, providing an optimal balance between accuracy and computational demand for these VQE calculations. Table \ref{tab:my_table16} adds an important layer of analysis by comparing our VQE and NumPy results to published data from the CCCBDB. For each system and basis set, the CCCBDB provides benchmark energy values against which we can evaluate the accuracy of our calculations. For example, the CCCBDB value for Al\textsubscript{2} with the STO-3G basis set is -477.145 74 \textit{E\textsubscript{h}}, which is close to our NumPy energy of -477.801 97 \textit{E\textsubscript{h}}. As we move to higher-level basis sets, such as 6-31G and cc-pVDZ, the VQE and NumPy results align more closely with the CCCBDB values, with errors consistently reducing. This validation against CCCBDB highlights that our VQE methodology, especially with the cc-pVDZ basis set, achieves high accuracy comparable to published data.

The comparison to CCCBDB data confirms that our VQE-based approach is robust and accurate for calculating ground state energies of aluminum systems. While minimal basis sets like STO-3G can provide rough estimates, using higher-level basis sets such as 6-31G and cc-pVDZ significantly improves the accuracy of our results, bringing them closer to established benchmark values. This analysis underscores the importance of basis set selection in quantum-DFT embedding workflows, especially for applications requiring high precision, as the choice of basis set directly impacts the reliability of energy estimations.
 
\begin{table}
\centering
\caption{Ground State Energy Results of Al\textsubscript{2} (\textit{E\textsubscript{h}}) on Statevector Simulator with Varied Basis Sets, using Default Parameters: SLSQP Optimizer, Efficient SU2 Ansatz (Circuit \#6), and 1 Repetition in Comparison to CCCBDB Values}
\label{tab:my_table16}
\scriptsize
\begin{tabular}{| l | l | l | l | l | l | l | l |}
\hline
Basis Set & VQE Energy & Variation (±)& Percent Error & NumPy Energy & Variation (±)& Percent Error & CCCBDB \\
\hline
STO-3G & -477.799 05& 0.000 000& 0.1369 \%& -477.801 97& 0.000 000& 0.1375 \%& -477.145 74\\
\hline
3-21G & -481.076 34& 0.012 279& 0.1153 \%& -481.095 08& 0.013 676& 0.1192 \%& -480.522 53\\
\hline
6-31G & -483.703 54& 0.000 732& 0.1133 \%& -483.706 66& 0.010 760& 0.1139 \%& -483.156 31\\
\hline
cc-PVDZ & -483.737 01& 0.011 890& 0.1119 \%& -483.741 61& 0.010 836& 0.1129 \%& -483.196 25\\
\hline

\end{tabular}

\end{table}

Lastly, to further evaluate the performance of the developed workflow under realistic conditions, we compared the VQE results for Al\textsubscript{2} obtained using noise models to the CCCBDB benchmark values (Table \ref{tab:comparison_energies}). Due to limited access to actual quantum hardware, we employed five IBM noise models—FakeSherbrooke, FakeManhattan, FakeToronto, FakeTokyo, and FakeMontreal—to simulate the effects of noise, such as gate errors, decoherence, and measurement inaccuracies. These noise models were employed on the QASM simulator because it mimics the probabilistic measurement process and incorporates realistic noise. According to IBM Quantum Documentation, noise models mimic the behaviors of quantum hardware by using snapshots, which contain critical information such as the coupling map, basis gates, and qubit properties. These details are crucial for performing noisy simulations that closely resemble real quantum hardware. \cite{qiskit2024} By providing an accurate depiction of real quantum devices, noise models are an essential tool for assessing quantum algorithms and workflows in the NISQ era.

\begin{table}
\centering
\caption{Ground State Energy Results of Al\textsubscript{2} (\textit{E\textsubscript{h}}) on the QASM Simulator including Noise Models, using Default Parameters: STO-3G Basis Set, SLSQP Optimizer, Efficient SU2 Ansatz (Circuit \#6), and 1 Repetition in Comparison to CCCBDB Values}
\label{tab:comparison_energies}
\begin{tabular}{| l  |l  |l  |l |}
\hline
Noise Model& VQE Energy & CCCBDB Energy & Percent Error \\
\hline
FakeSherbrooke & -477.211 77& -477.145 74& 0.01384 \%\\ \hline 
FakeManhattan & -477.173 84& -477.145 74& 0.00589 \%\\ \hline 
FakeToronto & -477.220 75& -477.145 74& 0.01572 \%\\ \hline 
FakeTokyo & -477.192 91& -477.145 74& 0.00989 \%\\ \hline
FakeMontreal & -477.203 17& -477.145 74& 0.01204 \%\\
\hline \hline
\end{tabular}
\end{table}

The results demonstrate close agreement with the CCCBDB reference energy of -477.145 74 \textit{E\textsubscript{h}}, with percent errors ranging from 0.005 89 \% (FakeManhattan) to 0.015 72 \% (FakeToronto). This consistency across noise models highlights the robustness of the VQE algorithm and the effectiveness of the quantum-DFT embedding workflow in producing reliable results under noisy conditions. Among the noise models, FakeManhattan exhibited the smallest deviation, suggesting that it most accurately replicates the performance of real hardware for this particular system. Conversely, FakeToronto showed the largest percent error, likely due to variations in its simulated error rates. Despite these differences, all results remained less than 0.02 \%, emphasizing the capability of the workflow to handle noise while delivering accurate energy estimates. This comparison validates the use of noise models when access to quantum hardware may be limited, particularly in the context of benchmarking algorithms. The agreement of the noise model results in comparison to the CCCBDB benchmark not only demonstrates the accuracy of the VQE algorithm but also reinforces the relevance of the developed workflow for quantum-enhanced chemistry simulations. These findings are especially promising for future applications, as they suggest that similar accuracy can be achieved on real quantum devices as hardware continues to improve.

The use of noise models also emphasizes the importance of benchmarking quantum algorithms under realistic conditions. As quantum hardware remains constrained by noise and limited qubit counts, noise-aware benchmarking provides critical insights into algorithm performance and helps identify areas for improvement. Future work could extend this analysis to other molecular systems in the CCCBDB to further validate the applicability of the workflow. Also, studying error mitigation techniques within noise models could help further narrow the gap between simulated and experimental results, enhancing the reliability and applicability of quantum computations in materials discovery.

\section{Conclusions and Future Perspectives}

This study demonstrates the efficacy of the Variational Quantum Eigensolver (VQE) within a quantum-DFT embedding framework for accurately simulating the ground state energies of aluminum clusters, with applications to materials discovery and design. By systematically benchmarking key parameters, including classical optimizers, circuit types, number of repetitions, simulator types, and basis sets, we achieved insights into optimizing VQE's performance on quantum simulators with and without noise. Notably, our findings show that SLSQP and COBYLA optimizers yield efficient convergence, while EfficientSU2 circuits and minimal repetitions provide reliable results without excessive computational cost. Furthermore, basis set selection proved critical; higher-level basis sets like cc-pVDZ enhanced accuracy, aligning our VQE results with NumPy benchmarks and published data from the CCCBDB. 

To simulate realistic conditions, we extended this benchmarking framework to include noise models, which emulate the behavior of actual quantum hardware. The results showed close agreement with the CCCBDB benchmark, with percent errors consistently below 0.02 \% across the five IBM noise models tested. The noise model results validate the robustness of the developed workflow and suggest that similar accuracy can be achieved on real quantum devices. Among the noise models, FakeManhattan demonstrated the smallest error. Our analysis highlights the capability of the VQE algorithm to produce accurate results even in the presence of noise, making it well-suited for applications in quantum-enhanced materials discovery.

Looking forward, our work opens avenues for expanded benchmarking and tool development to support quantum chemistry and materials science research. Future efforts will focus on extending this benchmarking framework to other systems in the CCCBDB, further validating the versatility of VQE for diverse chemical species. Additionally, we aim to integrate our findings into the JARVIS-AtomQC platform, making results readily accessible to the broader research community. Finally, we plan to develop a dedicated Python package that encapsulates our benchmarking workflow, enabling researchers to efficiently test and optimize quantum algorithms for material simulations. These advancements will contribute to establishing standardized practices in quantum algorithm benchmarking, accelerating progress toward practical quantum computing applications in materials discovery and beyond.

\section{Data Availability}
The code and data used in this work are available at \url{https://github.com/usnistgov/BenchQC}.


\section*{Acknowledgements}

The authors would also like to thank  the National Institute of Standards and Technology (NIST) for computational and funding support. 
Certain commercial equipment, instruments, software, or materials are identified in this paper in order to specify the experimental procedure adequately. Such identifications are not intended to imply recommendation or endorsement by NIST, nor it is intended to imply that the materials or equipment identified are necessarily the best available for the purpose.

\begin{suppinfo}

Please see Supporting Information for additional data tables including specific energy values for all systems. 

\end{suppinfo}

\bibliography{references}

\providecommand{\latin}[1]{#1}
\makeatletter
\providecommand{\doi}
  {\begingroup\let\do\@makeother\dospecials
  \catcode`\{=1 \catcode`\}=2 \doi@aux}
\providecommand{\doi@aux}[1]{\endgroup\texttt{#1}}
\makeatother
\providecommand*\mcitethebibliography{\thebibliography}
\csname @ifundefined\endcsname{endmcitethebibliography}  {\let\endmcitethebibliography\endthebibliography}{}
\begin{mcitethebibliography}{51}
\providecommand*\natexlab[1]{#1}
\providecommand*\mciteSetBstSublistMode[1]{}
\providecommand*\mciteSetBstMaxWidthForm[2]{}
\providecommand*\mciteBstWouldAddEndPuncttrue
  {\def\EndOfBibitem{\unskip.}}
\providecommand*\mciteBstWouldAddEndPunctfalse
  {\let\EndOfBibitem\relax}
\providecommand*\mciteSetBstMidEndSepPunct[3]{}
\providecommand*\mciteSetBstSublistLabelBeginEnd[3]{}
\providecommand*\EndOfBibitem{}
\mciteSetBstSublistMode{f}
\mciteSetBstMaxWidthForm{subitem}{(\alph{mcitesubitemcount})}
\mciteSetBstSublistLabelBeginEnd
  {\mcitemaxwidthsubitemform\space}
  {\relax}
  {\relax}

\bibitem[Nielsen and Chuang(2010)Nielsen, and Chuang]{Nielsen_Chuang_2010}
Nielsen,~M.~A.; Chuang,~I.~L. \emph{Quantum Computation and Quantum Information: 10th Anniversary Edition}; Cambridge University Press, 2010\relax
\mciteBstWouldAddEndPuncttrue
\mciteSetBstMidEndSepPunct{\mcitedefaultmidpunct}
{\mcitedefaultendpunct}{\mcitedefaultseppunct}\relax
\EndOfBibitem
\bibitem[Preskill(2018)]{Preskill2018quantumcomputingin}
Preskill,~J. Quantum {C}omputing in the {NISQ} era and beyond. \emph{{Quantum}} \textbf{2018}, \emph{2}, 79\relax
\mciteBstWouldAddEndPuncttrue
\mciteSetBstMidEndSepPunct{\mcitedefaultmidpunct}
{\mcitedefaultendpunct}{\mcitedefaultseppunct}\relax
\EndOfBibitem
\bibitem[Arute \latin{et~al.}(2019)Arute, Arya, Babbush, Bacon, Bardin, Barends, Biswas, Boixo, Brandao, Buell, \latin{et~al.} others]{arute2019quantum}
Arute,~F.; Arya,~K.; Babbush,~R.; Bacon,~D.; Bardin,~J.~C.; Barends,~R.; Biswas,~R.; Boixo,~S.; Brandao,~F.~G.; Buell,~D.~A.; others Quantum supremacy using a programmable superconducting processor. \emph{Nature} \textbf{2019}, \emph{574}, 505--510\relax
\mciteBstWouldAddEndPuncttrue
\mciteSetBstMidEndSepPunct{\mcitedefaultmidpunct}
{\mcitedefaultendpunct}{\mcitedefaultseppunct}\relax
\EndOfBibitem
\bibitem[Bravyi \latin{et~al.}(2018)Bravyi, Gosset, and K{\"o}nig]{bravyi2018quantum}
Bravyi,~S.; Gosset,~D.; K{\"o}nig,~R. Quantum advantage with shallow circuits. \emph{Science} \textbf{2018}, \emph{362}, 308--311\relax
\mciteBstWouldAddEndPuncttrue
\mciteSetBstMidEndSepPunct{\mcitedefaultmidpunct}
{\mcitedefaultendpunct}{\mcitedefaultseppunct}\relax
\EndOfBibitem
\bibitem[Aspuru-Guzik \latin{et~al.}(2005)Aspuru-Guzik, Dutoi, Love, and Head-Gordon]{aspuru2005simulated}
Aspuru-Guzik,~A.; Dutoi,~A.~D.; Love,~P.~J.; Head-Gordon,~M. Simulated quantum computation of molecular energies. \emph{Science} \textbf{2005}, \emph{309}, 1704--1707\relax
\mciteBstWouldAddEndPuncttrue
\mciteSetBstMidEndSepPunct{\mcitedefaultmidpunct}
{\mcitedefaultendpunct}{\mcitedefaultseppunct}\relax
\EndOfBibitem
\bibitem[McArdle \latin{et~al.}(2020)McArdle, Endo, Aspuru-Guzik, Benjamin, and Yuan]{mcardle2020quantum}
McArdle,~S.; Endo,~S.; Aspuru-Guzik,~A.; Benjamin,~S.~C.; Yuan,~X. Quantum computational chemistry. \emph{Reviews of Modern Physics} \textbf{2020}, \emph{92}, 015003\relax
\mciteBstWouldAddEndPuncttrue
\mciteSetBstMidEndSepPunct{\mcitedefaultmidpunct}
{\mcitedefaultendpunct}{\mcitedefaultseppunct}\relax
\EndOfBibitem
\bibitem[Bauer \latin{et~al.}(2020)Bauer, Bravyi, Motta, and Chan]{bauer2020quantum}
Bauer,~B.; Bravyi,~S.; Motta,~M.; Chan,~G. K.-L. Quantum algorithms for quantum chemistry and quantum materials science. \emph{Chemical Reviews} \textbf{2020}, \emph{120}, 12685--12717\relax
\mciteBstWouldAddEndPuncttrue
\mciteSetBstMidEndSepPunct{\mcitedefaultmidpunct}
{\mcitedefaultendpunct}{\mcitedefaultseppunct}\relax
\EndOfBibitem
\bibitem[Booth \latin{et~al.}(2013)Booth, Gr{\"u}neis, Kresse, and Alavi]{booth2013towards}
Booth,~G.~H.; Gr{\"u}neis,~A.; Kresse,~G.; Alavi,~A. Towards an exact description of electronic wavefunctions in real solids. \emph{Nature} \textbf{2013}, \emph{493}, 365--370\relax
\mciteBstWouldAddEndPuncttrue
\mciteSetBstMidEndSepPunct{\mcitedefaultmidpunct}
{\mcitedefaultendpunct}{\mcitedefaultseppunct}\relax
\EndOfBibitem
\bibitem[Motta \latin{et~al.}(2017)Motta, Ceperley, Chan, Gomez, Gull, Guo, Jim{\'e}nez-Hoyos, Lan, Li, Ma, \latin{et~al.} others]{motta2017towards}
Motta,~M.; Ceperley,~D.~M.; Chan,~G. K.-L.; Gomez,~J.~A.; Gull,~E.; Guo,~S.; Jim{\'e}nez-Hoyos,~C.~A.; Lan,~T.~N.; Li,~J.; Ma,~F.; others Towards the solution of the many-electron problem in real materials: Equation of state of the hydrogen chain with state-of-the-art many-body methods. \emph{Physical Review X} \textbf{2017}, \emph{7}, 031059\relax
\mciteBstWouldAddEndPuncttrue
\mciteSetBstMidEndSepPunct{\mcitedefaultmidpunct}
{\mcitedefaultendpunct}{\mcitedefaultseppunct}\relax
\EndOfBibitem
\bibitem[Cohen \latin{et~al.}(2012)Cohen, Mori-S{\'a}nchez, and Yang]{cohen2012challenges}
Cohen,~A.~J.; Mori-S{\'a}nchez,~P.; Yang,~W. Challenges for density functional theory. \emph{Chemical reviews} \textbf{2012}, \emph{112}, 289--320\relax
\mciteBstWouldAddEndPuncttrue
\mciteSetBstMidEndSepPunct{\mcitedefaultmidpunct}
{\mcitedefaultendpunct}{\mcitedefaultseppunct}\relax
\EndOfBibitem
\bibitem[Alexeev \latin{et~al.}(2024)Alexeev, Amsler, Barroca, Bassini, Battelle, Camps, Casanova, Choi, Chong, Chung, \latin{et~al.} others]{alexeev2024quantum}
Alexeev,~Y.; Amsler,~M.; Barroca,~M.~A.; Bassini,~S.; Battelle,~T.; Camps,~D.; Casanova,~D.; Choi,~Y.~J.; Chong,~F.~T.; Chung,~C.; others Quantum-centric supercomputing for materials science: A perspective on challenges and future directions. \emph{Future Generation Computer Systems} \textbf{2024}, \emph{160}, 666--710\relax
\mciteBstWouldAddEndPuncttrue
\mciteSetBstMidEndSepPunct{\mcitedefaultmidpunct}
{\mcitedefaultendpunct}{\mcitedefaultseppunct}\relax
\EndOfBibitem
\bibitem[Rajan(2015)]{rajan2015materials}
Rajan,~K. Materials informatics: The materials “gene” and big data. \emph{Annual Review of Materials Research} \textbf{2015}, \emph{45}, 153--169\relax
\mciteBstWouldAddEndPuncttrue
\mciteSetBstMidEndSepPunct{\mcitedefaultmidpunct}
{\mcitedefaultendpunct}{\mcitedefaultseppunct}\relax
\EndOfBibitem
\bibitem[Bharti \latin{et~al.}(2022)Bharti, Cervera-Lierta, Kyaw, Haug, Alperin-Lea, Anand, Degroote, Heimonen, Kottmann, Menke, \latin{et~al.} others]{bharti2022noisy}
Bharti,~K.; Cervera-Lierta,~A.; Kyaw,~T.~H.; Haug,~T.; Alperin-Lea,~S.; Anand,~A.; Degroote,~M.; Heimonen,~H.; Kottmann,~J.~S.; Menke,~T.; others Noisy intermediate-scale quantum algorithms. \emph{Reviews of Modern Physics} \textbf{2022}, \emph{94}, 015004\relax
\mciteBstWouldAddEndPuncttrue
\mciteSetBstMidEndSepPunct{\mcitedefaultmidpunct}
{\mcitedefaultendpunct}{\mcitedefaultseppunct}\relax
\EndOfBibitem
\bibitem[Rossmannek \latin{et~al.}(2021)Rossmannek, Barkoutsos, Ollitrault, and Tavernelli]{rossmannek2021quantum}
Rossmannek,~M.; Barkoutsos,~P.~K.; Ollitrault,~P.~J.; Tavernelli,~I. Quantum HF/DFT-embedding algorithms for electronic structure calculations: Scaling up to complex molecular systems. \emph{The Journal of Chemical Physics} \textbf{2021}, \emph{154}\relax
\mciteBstWouldAddEndPuncttrue
\mciteSetBstMidEndSepPunct{\mcitedefaultmidpunct}
{\mcitedefaultendpunct}{\mcitedefaultseppunct}\relax
\EndOfBibitem
\bibitem[Parrish \latin{et~al.}(2019)Parrish, Hohenstein, McMahon, and Mart{\'\i}nez]{parrish2019quantum}
Parrish,~R.~M.; Hohenstein,~E.~G.; McMahon,~P.~L.; Mart{\'\i}nez,~T.~J. Quantum computation of electronic transitions using a variational quantum eigensolver. \emph{Physical review letters} \textbf{2019}, \emph{122}, 230401\relax
\mciteBstWouldAddEndPuncttrue
\mciteSetBstMidEndSepPunct{\mcitedefaultmidpunct}
{\mcitedefaultendpunct}{\mcitedefaultseppunct}\relax
\EndOfBibitem
\bibitem[Rossmannek \latin{et~al.}(2023)Rossmannek, Pavosevic, Rubio, and Tavernelli]{rossmannek2023quantum}
Rossmannek,~M.; Pavosevic,~F.; Rubio,~A.; Tavernelli,~I. Quantum embedding method for the simulation of strongly correlated systems on quantum computers. \emph{The Journal of Physical Chemistry Letters} \textbf{2023}, \emph{14}, 3491--3497\relax
\mciteBstWouldAddEndPuncttrue
\mciteSetBstMidEndSepPunct{\mcitedefaultmidpunct}
{\mcitedefaultendpunct}{\mcitedefaultseppunct}\relax
\EndOfBibitem
\bibitem[Peruzzo \latin{et~al.}(2014)Peruzzo, McClean, Shadbolt, Yung, Zhou, Love, Aspuru-Guzik, and O’brien]{peruzzo2014variational}
Peruzzo,~A.; McClean,~J.; Shadbolt,~P.; Yung,~M.-H.; Zhou,~X.-Q.; Love,~P.~J.; Aspuru-Guzik,~A.; O’brien,~J.~L. A variational eigenvalue solver on a photonic quantum processor. \emph{Nature communications} \textbf{2014}, \emph{5}, 4213\relax
\mciteBstWouldAddEndPuncttrue
\mciteSetBstMidEndSepPunct{\mcitedefaultmidpunct}
{\mcitedefaultendpunct}{\mcitedefaultseppunct}\relax
\EndOfBibitem
\bibitem[McClean \latin{et~al.}(2016)McClean, Romero, Babbush, and Aspuru-Guzik]{mcclean2016theory}
McClean,~J.~R.; Romero,~J.; Babbush,~R.; Aspuru-Guzik,~A. The theory of variational hybrid quantum-classical algorithms. \emph{New Journal of Physics} \textbf{2016}, \emph{18}, 023023\relax
\mciteBstWouldAddEndPuncttrue
\mciteSetBstMidEndSepPunct{\mcitedefaultmidpunct}
{\mcitedefaultendpunct}{\mcitedefaultseppunct}\relax
\EndOfBibitem
\bibitem[Cao \latin{et~al.}(2019)Cao, Romero, Olson, Degroote, Johnson, Kieferov{\'a}, Kivlichan, Menke, Peropadre, Sawaya, \latin{et~al.} others]{cao2019quantum}
Cao,~Y.; Romero,~J.; Olson,~J.~P.; Degroote,~M.; Johnson,~P.~D.; Kieferov{\'a},~M.; Kivlichan,~I.~D.; Menke,~T.; Peropadre,~B.; Sawaya,~N.~P.; others Quantum chemistry in the age of quantum computing. \emph{Chemical reviews} \textbf{2019}, \emph{119}, 10856--10915\relax
\mciteBstWouldAddEndPuncttrue
\mciteSetBstMidEndSepPunct{\mcitedefaultmidpunct}
{\mcitedefaultendpunct}{\mcitedefaultseppunct}\relax
\EndOfBibitem
\bibitem[Kandala \latin{et~al.}(2017)Kandala, Mezzacapo, Temme, Takita, Brink, Chow, and Gambetta]{kandala2017hardware}
Kandala,~A.; Mezzacapo,~A.; Temme,~K.; Takita,~M.; Brink,~M.; Chow,~J.~M.; Gambetta,~J.~M. Hardware-efficient variational quantum eigensolver for small molecules and quantum magnets. \emph{nature} \textbf{2017}, \emph{549}, 242--246\relax
\mciteBstWouldAddEndPuncttrue
\mciteSetBstMidEndSepPunct{\mcitedefaultmidpunct}
{\mcitedefaultendpunct}{\mcitedefaultseppunct}\relax
\EndOfBibitem
\bibitem[Fedorov \latin{et~al.}(2022)Fedorov, Peng, Govind, and Alexeev]{fedorov2022vqe}
Fedorov,~D.~A.; Peng,~B.; Govind,~N.; Alexeev,~Y. VQE method: a short survey and recent developments. \emph{Materials Theory} \textbf{2022}, \emph{6}, 2\relax
\mciteBstWouldAddEndPuncttrue
\mciteSetBstMidEndSepPunct{\mcitedefaultmidpunct}
{\mcitedefaultendpunct}{\mcitedefaultseppunct}\relax
\EndOfBibitem
\bibitem[Rubin \latin{et~al.}(2018)Rubin, Babbush, and McClean]{rubin2018application}
Rubin,~N.~C.; Babbush,~R.; McClean,~J. Application of fermionic marginal constraints to hybrid quantum algorithms. \emph{New Journal of Physics} \textbf{2018}, \emph{20}, 053020\relax
\mciteBstWouldAddEndPuncttrue
\mciteSetBstMidEndSepPunct{\mcitedefaultmidpunct}
{\mcitedefaultendpunct}{\mcitedefaultseppunct}\relax
\EndOfBibitem
\bibitem[O’Malley \latin{et~al.}(2016)O’Malley, Babbush, Kivlichan, Romero, McClean, Barends, Kelly, Roushan, Tranter, Ding, \latin{et~al.} others]{o2016scalable}
O’Malley,~P.~J.; Babbush,~R.; Kivlichan,~I.~D.; Romero,~J.; McClean,~J.~R.; Barends,~R.; Kelly,~J.; Roushan,~P.; Tranter,~A.; Ding,~N.; others Scalable quantum simulation of molecular energies. \emph{Physical Review X} \textbf{2016}, \emph{6}, 031007\relax
\mciteBstWouldAddEndPuncttrue
\mciteSetBstMidEndSepPunct{\mcitedefaultmidpunct}
{\mcitedefaultendpunct}{\mcitedefaultseppunct}\relax
\EndOfBibitem
\bibitem[Hu \latin{et~al.}(2022)Hu, Li, Lin, Long, Xu, Su, Zhang, Zhu, and Yung]{hu2022benchmarkingvariationalquantumeigensolvers}
Hu,~J.; Li,~J.; Lin,~Y.; Long,~H.; Xu,~X.-S.; Su,~Z.; Zhang,~W.; Zhu,~Y.; Yung,~M.-H. Benchmarking Variational Quantum Eigensolvers for Quantum Chemistry. 2022; \url{https://arxiv.org/abs/2211.12775}\relax
\mciteBstWouldAddEndPuncttrue
\mciteSetBstMidEndSepPunct{\mcitedefaultmidpunct}
{\mcitedefaultendpunct}{\mcitedefaultseppunct}\relax
\EndOfBibitem
\bibitem[Singh \latin{et~al.}(2023)Singh, Majumder, and Mishra]{10.1063/5.0161057}
Singh,~H.; Majumder,~S.; Mishra,~S. Benchmarking of different optimizers in the variational quantum algorithms for applications in quantum chemistry. \emph{The Journal of Chemical Physics} \textbf{2023}, \emph{159}, 044117\relax
\mciteBstWouldAddEndPuncttrue
\mciteSetBstMidEndSepPunct{\mcitedefaultmidpunct}
{\mcitedefaultendpunct}{\mcitedefaultseppunct}\relax
\EndOfBibitem
\bibitem[Choudhary(2021)]{choudhary2021quantum}
Choudhary,~K. Quantum computation for predicting electron and phonon properties of solids. \emph{Journal of Physics: Condensed Matter} \textbf{2021}, \emph{33}, 385501\relax
\mciteBstWouldAddEndPuncttrue
\mciteSetBstMidEndSepPunct{\mcitedefaultmidpunct}
{\mcitedefaultendpunct}{\mcitedefaultseppunct}\relax
\EndOfBibitem
\bibitem[{Sivakumar, Ashwin} \latin{et~al.}(2024){Sivakumar, Ashwin}, {K Nair, Harishankar}, {Joshi, Aurum}, {R, Kenson Wesley}, {P Videsh, Akash}, and {P, Reena Monica}]{refId0}
{Sivakumar, Ashwin}; {K Nair, Harishankar}; {Joshi, Aurum}; {R, Kenson Wesley}; {P Videsh, Akash}; {P, Reena Monica} A computational study and analysis of Variational Quantum Eigensolver over multiple parameters for molecules and ions. \emph{EPJ Quantum Technol.} \textbf{2024}, \emph{11}, 73\relax
\mciteBstWouldAddEndPuncttrue
\mciteSetBstMidEndSepPunct{\mcitedefaultmidpunct}
{\mcitedefaultendpunct}{\mcitedefaultseppunct}\relax
\EndOfBibitem
\bibitem[Pollard and Clayborne(2025)Pollard, and Clayborne]{pollard2025quantum}
Pollard,~N.; Clayborne,~A. Manuscript in preparation\relax
\mciteBstWouldAddEndPuncttrue
\mciteSetBstMidEndSepPunct{\mcitedefaultmidpunct}
{\mcitedefaultendpunct}{\mcitedefaultseppunct}\relax
\EndOfBibitem
\bibitem[Javadi-Abhari \latin{et~al.}(2024)Javadi-Abhari, Treinish, Krsulich, Wood, Lishman, Gacon, Martiel, Nation, Bishop, Cross, Johnson, and Gambetta]{qiskit2024}
Javadi-Abhari,~A.; Treinish,~M.; Krsulich,~K.; Wood,~C.~J.; Lishman,~J.; Gacon,~J.; Martiel,~S.; Nation,~P.~D.; Bishop,~L.~S.; Cross,~A.~W.; Johnson,~B.~R.; Gambetta,~J.~M. Quantum computing with {Q}iskit. 2024\relax
\mciteBstWouldAddEndPuncttrue
\mciteSetBstMidEndSepPunct{\mcitedefaultmidpunct}
{\mcitedefaultendpunct}{\mcitedefaultseppunct}\relax
\EndOfBibitem
\bibitem[IIIC(2022)]{iiic2022nist}
IIIC,~R.~J. NIST Computational Chemistry Comparison and Benchmark Database. 2022\relax
\mciteBstWouldAddEndPuncttrue
\mciteSetBstMidEndSepPunct{\mcitedefaultmidpunct}
{\mcitedefaultendpunct}{\mcitedefaultseppunct}\relax
\EndOfBibitem
\bibitem[Choudhary \latin{et~al.}(2020)Choudhary, Garrity, Reid, DeCost, Biacchi, Hight~Walker, Trautt, Hattrick-Simpers, Kusne, Centrone, \latin{et~al.} others]{choudhary2020joint}
Choudhary,~K.; Garrity,~K.~F.; Reid,~A.~C.; DeCost,~B.; Biacchi,~A.~J.; Hight~Walker,~A.~R.; Trautt,~Z.; Hattrick-Simpers,~J.; Kusne,~A.~G.; Centrone,~A.; others The joint automated repository for various integrated simulations (JARVIS) for data-driven materials design. \emph{npj computational materials} \textbf{2020}, \emph{6}, 173\relax
\mciteBstWouldAddEndPuncttrue
\mciteSetBstMidEndSepPunct{\mcitedefaultmidpunct}
{\mcitedefaultendpunct}{\mcitedefaultseppunct}\relax
\EndOfBibitem
\bibitem[Wines \latin{et~al.}(2023)Wines, Gurunathan, Garrity, DeCost, Biacchi, Tavazza, and Choudhary]{10.1063/5.0159299}
Wines,~D.; Gurunathan,~R.; Garrity,~K.~F.; DeCost,~B.; Biacchi,~A.~J.; Tavazza,~F.; Choudhary,~K. Recent progress in the JARVIS infrastructure for next-generation data-driven materials design. \emph{Applied Physics Reviews} \textbf{2023}, \emph{10}, 041302\relax
\mciteBstWouldAddEndPuncttrue
\mciteSetBstMidEndSepPunct{\mcitedefaultmidpunct}
{\mcitedefaultendpunct}{\mcitedefaultseppunct}\relax
\EndOfBibitem
\bibitem[Sun \latin{et~al.}(2020)Sun, Zhang, Banerjee, Bao, Barbry, Blunt, Bogdanov, Booth, Chen, Cui, Eriksen, Gao, Guo, Hermann, Hermes, Koh, Koval, Lehtola, Li, Liu, Mardirossian, McClain, Motta, Mussard, Pham, Pulkin, Purwanto, Robinson, Ronca, Sayfutyarova, Scheurer, Schurkus, Smith, Sun, Sun, Upadhyay, Wagner, Wang, White, Whitfield, Williamson, Wouters, Yang, Yu, Zhu, Berkelbach, Sharma, Sokolov, and Chan]{10.1063/5.0006074}
Sun,~Q. \latin{et~al.}  {Recent developments in the PySCF program package}. \emph{The Journal of Chemical Physics} \textbf{2020}, \emph{153}, 024109\relax
\mciteBstWouldAddEndPuncttrue
\mciteSetBstMidEndSepPunct{\mcitedefaultmidpunct}
{\mcitedefaultendpunct}{\mcitedefaultseppunct}\relax
\EndOfBibitem
\bibitem[Sun \latin{et~al.}(2018)Sun, Berkelbach, Blunt, Booth, Guo, Li, Liu, McClain, Sayfutyarova, Sharma, \latin{et~al.} others]{sun2018pyscf}
Sun,~Q.; Berkelbach,~T.~C.; Blunt,~N.~S.; Booth,~G.~H.; Guo,~S.; Li,~Z.; Liu,~J.; McClain,~J.~D.; Sayfutyarova,~E.~R.; Sharma,~S.; others PySCF: the Python-based simulations of chemistry framework. \emph{Wiley Interdisciplinary Reviews: Computational Molecular Science} \textbf{2018}, \emph{8}, e1340\relax
\mciteBstWouldAddEndPuncttrue
\mciteSetBstMidEndSepPunct{\mcitedefaultmidpunct}
{\mcitedefaultendpunct}{\mcitedefaultseppunct}\relax
\EndOfBibitem
\bibitem[Sun(2015)]{sun2015libcint}
Sun,~Q. Libcint: An efficient general integral library for g aussian basis functions. \emph{Journal of computational chemistry} \textbf{2015}, \emph{36}, 1664--1671\relax
\mciteBstWouldAddEndPuncttrue
\mciteSetBstMidEndSepPunct{\mcitedefaultmidpunct}
{\mcitedefaultendpunct}{\mcitedefaultseppunct}\relax
\EndOfBibitem
\bibitem[{Qiskit Nature Developers}(2021)]{qiskit_nature_2021}
{Qiskit Nature Developers} Qiskit Nature: A library for quantum computing in chemistry and physics. 2021; \url{https://qiskit.org/documentation/nature/}\relax
\mciteBstWouldAddEndPuncttrue
\mciteSetBstMidEndSepPunct{\mcitedefaultmidpunct}
{\mcitedefaultendpunct}{\mcitedefaultseppunct}\relax
\EndOfBibitem
\bibitem[Mason(2015)]{mason2015aluminum}
Mason,~M.~R. \emph{Aluminum-Based Catalysis}; John Wiley \& Sons, Ltd: New York, 2015\relax
\mciteBstWouldAddEndPuncttrue
\mciteSetBstMidEndSepPunct{\mcitedefaultmidpunct}
{\mcitedefaultendpunct}{\mcitedefaultseppunct}\relax
\EndOfBibitem
\bibitem[Graves \latin{et~al.}(2005)Graves, Campbell, and Nguyen]{graves2005aluminum}
Graves,~C.~R.; Campbell,~E.~J.; Nguyen,~S.~T. Aluminum-based catalysts for the asymmetric Meerwein--Schmidt--Ponndorf--Verley--Oppenauer (MSPVO) reaction manifold. \emph{Tetrahedron: Asymmetry} \textbf{2005}, \emph{16}, 3460--3468\relax
\mciteBstWouldAddEndPuncttrue
\mciteSetBstMidEndSepPunct{\mcitedefaultmidpunct}
{\mcitedefaultendpunct}{\mcitedefaultseppunct}\relax
\EndOfBibitem
\bibitem[Arakawa \latin{et~al.}(2015)Arakawa, Kohara, and Terasaki]{arakawa2015reaction}
Arakawa,~M.; Kohara,~K.; Terasaki,~A. Reaction of aluminum cluster cations with a mixture of O2 and H2O gases: Formation of hydrated-alumina clusters. \emph{The Journal of Physical Chemistry C} \textbf{2015}, \emph{119}, 10981--10986\relax
\mciteBstWouldAddEndPuncttrue
\mciteSetBstMidEndSepPunct{\mcitedefaultmidpunct}
{\mcitedefaultendpunct}{\mcitedefaultseppunct}\relax
\EndOfBibitem
\bibitem[Kohn and Sham(1965)Kohn, and Sham]{kohn1965self}
Kohn,~W.; Sham,~L.~J. Self-consistent equations including exchange and correlation effects. \emph{Physical review} \textbf{1965}, \emph{140}, A1133\relax
\mciteBstWouldAddEndPuncttrue
\mciteSetBstMidEndSepPunct{\mcitedefaultmidpunct}
{\mcitedefaultendpunct}{\mcitedefaultseppunct}\relax
\EndOfBibitem
\bibitem[Momma and Izumi(2011)Momma, and Izumi]{momma2011vesta}
Momma,~K.; Izumi,~F. VESTA 3 for three-dimensional visualization of crystal, volumetric and morphology data. \emph{Journal of applied crystallography} \textbf{2011}, \emph{44}, 1272--1276\relax
\mciteBstWouldAddEndPuncttrue
\mciteSetBstMidEndSepPunct{\mcitedefaultmidpunct}
{\mcitedefaultendpunct}{\mcitedefaultseppunct}\relax
\EndOfBibitem
\bibitem[Simons(2008)]{simons2008molecular}
Simons,~J. Molecular anions. \emph{The Journal of Physical Chemistry A} \textbf{2008}, \emph{112}, 6401--6511\relax
\mciteBstWouldAddEndPuncttrue
\mciteSetBstMidEndSepPunct{\mcitedefaultmidpunct}
{\mcitedefaultendpunct}{\mcitedefaultseppunct}\relax
\EndOfBibitem
\bibitem[Jensen(2017)]{jensen2017introduction}
Jensen,~F. \emph{Introduction to computational chemistry}; John wiley \& sons, 2017\relax
\mciteBstWouldAddEndPuncttrue
\mciteSetBstMidEndSepPunct{\mcitedefaultmidpunct}
{\mcitedefaultendpunct}{\mcitedefaultseppunct}\relax
\EndOfBibitem
\bibitem[Bag \latin{et~al.}(2017)Bag, Porzelt, Altmann, and Inoue]{bag2017stable}
Bag,~P.; Porzelt,~A.; Altmann,~P.~J.; Inoue,~S. A stable neutral compound with an aluminum--aluminum double bond. \emph{Journal of the American Chemical Society} \textbf{2017}, \emph{139}, 14384--14387\relax
\mciteBstWouldAddEndPuncttrue
\mciteSetBstMidEndSepPunct{\mcitedefaultmidpunct}
{\mcitedefaultendpunct}{\mcitedefaultseppunct}\relax
\EndOfBibitem
\bibitem[Lavrijsen \latin{et~al.}(2020)Lavrijsen, Tudor, M{\"u}ller, Iancu, and De~Jong]{lavrijsen2020classical}
Lavrijsen,~W.; Tudor,~A.; M{\"u}ller,~J.; Iancu,~C.; De~Jong,~W. Classical optimizers for noisy intermediate-scale quantum devices. 2020 IEEE international conference on quantum computing and engineering (QCE). 2020; pp 267--277\relax
\mciteBstWouldAddEndPuncttrue
\mciteSetBstMidEndSepPunct{\mcitedefaultmidpunct}
{\mcitedefaultendpunct}{\mcitedefaultseppunct}\relax
\EndOfBibitem
\bibitem[Pellow-Jarman \latin{et~al.}(2021)Pellow-Jarman, Sinayskiy, Pillay, and Petruccione]{pellow2021comparison}
Pellow-Jarman,~A.; Sinayskiy,~I.; Pillay,~A.; Petruccione,~F. A comparison of various classical optimizers for a variational quantum linear solver. \emph{Quantum Information Processing} \textbf{2021}, \emph{20}, 202\relax
\mciteBstWouldAddEndPuncttrue
\mciteSetBstMidEndSepPunct{\mcitedefaultmidpunct}
{\mcitedefaultendpunct}{\mcitedefaultseppunct}\relax
\EndOfBibitem
\bibitem[Sim \latin{et~al.}(2019)Sim, Johnson, and Aspuru-Guzik]{sim2019expressibility}
Sim,~S.; Johnson,~P.~D.; Aspuru-Guzik,~A. Expressibility and entangling capability of parameterized quantum circuits for hybrid quantum-classical algorithms. \emph{Advanced Quantum Technologies} \textbf{2019}, \emph{2}, 1900070\relax
\mciteBstWouldAddEndPuncttrue
\mciteSetBstMidEndSepPunct{\mcitedefaultmidpunct}
{\mcitedefaultendpunct}{\mcitedefaultseppunct}\relax
\EndOfBibitem
\bibitem[Holmes \latin{et~al.}(2022)Holmes, Sharma, Cerezo, and Coles]{holmes2022connecting}
Holmes,~Z.; Sharma,~K.; Cerezo,~M.; Coles,~P.~J. Connecting ansatz expressibility to gradient magnitudes and barren plateaus. \emph{PRX Quantum} \textbf{2022}, \emph{3}, 010313\relax
\mciteBstWouldAddEndPuncttrue
\mciteSetBstMidEndSepPunct{\mcitedefaultmidpunct}
{\mcitedefaultendpunct}{\mcitedefaultseppunct}\relax
\EndOfBibitem
\bibitem[Aleksandrowicz \latin{et~al.}(2019)Aleksandrowicz, Alexander, Barkoutsos, Bello, Ben-Haim, Bucher, Cabrera-Hern{\'a}ndez, Carballo-Franquis, Chen, Chen, \latin{et~al.} others]{aleksandrowicz2019qiskit}
Aleksandrowicz,~G.; Alexander,~T.; Barkoutsos,~P.; Bello,~L.; Ben-Haim,~Y.; Bucher,~D.; Cabrera-Hern{\'a}ndez,~F.~J.; Carballo-Franquis,~J.; Chen,~A.; Chen,~C.-F.; others Qiskit: An open-source framework for quantum computing. \emph{Accessed on: Mar} \textbf{2019}, \emph{16}, 61\relax
\mciteBstWouldAddEndPuncttrue
\mciteSetBstMidEndSepPunct{\mcitedefaultmidpunct}
{\mcitedefaultendpunct}{\mcitedefaultseppunct}\relax
\EndOfBibitem
\bibitem[ccc(2022)]{cccbdb}
{NIST Computational Chemistry Comparison and Benchmark Database}. \url{http://cccbdb.nist.gov/}, 2022; NIST Standard Reference Database Number 101, Release 22, May 2022\relax
\mciteBstWouldAddEndPuncttrue
\mciteSetBstMidEndSepPunct{\mcitedefaultmidpunct}
{\mcitedefaultendpunct}{\mcitedefaultseppunct}\relax
\EndOfBibitem
\end{mcitethebibliography}

\end{document}









Default parameters of all simulations: STO-3G Basis Set, the Sequential Least Squares Programming (SLSQP) Optimizer, an Efficient SU2 Ansatz (Circuit \#6), and 1 Repetition on the Statevector Simulator

\begin{table}
\centering
\caption{Ground State Energy (\textit{E\textsubscript{h}}) Results of Varying Classical Optimizers (SLSQP, Limited-memory Broyden–Fletcher–Goldfarb–Shanno Bound (L\_BFGS\_B), Constrained Optimization By Linear Approximations (COBYLA), and Simultaneous Perturbation Stochastic Approximation (SPSA) on Statevector Simulator (Al\textsuperscript{- })}
\label{tab:my_table1}
\begin{tabular}{| l  |l  |l  |l  |}
\hline
 & Al\textsuperscript{- }(VQE)& Variation (±)& Al\textsuperscript{- }(NumPy)\\
\hline
SLSQP & -238.842 31& 0.000 000& -238.870 47\\
\hline
L\_BFGS\_B & -238.842 29& 0.000 000& -238.870 47\\
\hline
COBYLA & -238.843 67& 0.000 037& -238.870 47\\
\hline
SPSA & -238.843 47& 0.000 000& -238.870 47\\
\hline

\end{tabular}

\end{table}
\begin{table}
\centering
\caption{Ground State Energy (\textit{E\textsubscript{h}}) Results of Varying Classical Optimizers (SLSQP, L\_BFGS\_B, COBYLA, SPSA) on Statevector Simulator (Al\textsubscript{2})}
\label{tab:my_table2}
\begin{tabular}{| l  |l  |l  |l  |}
\hline
 & Al\textsubscript{2} (VQE) & Variation (±)& Al\textsubscript{2} (NumPy) \\
\hline
SLSQP & -477.799 05& 0.000 000&  -477.801 97\\
\hline
L\_BFGS\_B & -477.799 05& 0.000 000& -477.801 97\\
\hline
COBYLA & -477.799 03& 0.000 022& -477.801 97\\
\hline
SPSA & -477.793 62& 0.000 000& -477.801 97\\
\hline

\end{tabular}

\end{table}

\begin{table}
\centering
\caption{Ground State Energy (\textit{E\textsubscript{h}}) Results of Varying Classical Optimizers (SLSQP, L\_BFGS\_B, COBYLA, SPSA) on Statevector Simulator (Al\textsubscript{3}\textsuperscript{- })}
\label{tab:my_table3}
\begin{tabular}{| l  |l  |l  |l  |}
\hline
 & Al\textsubscript{3}\textsuperscript{- }(VQE) & Variation (±)& Al\textsubscript{3}\textsuperscript{- }(NumPy) \\
\hline
SLSQP & -716.675 25& 0.000 000& -716.708 73\\
\hline
L\_BFGS\_B & -716.679 28& 0.000 031& -716.708 73\\
\hline
COBYLA & -716.690 60& 0.000 000& -716.708 73\\
\hline
SPSA & -716.698 41& 0.000 000& -716.708 73\\
\hline

\end{tabular}

\end{table}

\begin{table}
\centering
\caption{Ground State Energy (\textit{E\textsubscript{h}}) Results of Varying Circuit Type on Statevector Simulator (Al\textsuperscript{- })}
\label{tab:table 4}
\begin{tabular}{| l  |l  |l  |l  |}
\hline
 & Al\textsuperscript{- }(VQE)& Variation (±)& Al\textsuperscript{- }(NumPy)\\
\hline
Circuit \#1 & -238.842 45& 0.000 000& -238.870 47\\
\hline
Circuit \#2 & -238.842 42& 0.000 000& -238.870 47\\
\hline
Circuit \#3 & -238.842 42& 0.000 000& -238.870 47\\
\hline
Circuit \#4 & -238.842 39& 0.000 000& -238.870 47\\
\hline
Circuit \#5 & -238.722 95& 0.061 264& -238.870 47\\
\hline
Circuit \#6 & -238.842 31& 0.000 000& -238.870 47\\
\hline

\end{tabular}

\end{table}

\begin{table}
\centering
\caption{Ground State Energy (\textit{E\textsubscript{h}}) Results of Varying Circuit Type on Statevector Simulator (Al\textsubscript{2})}
\label{tab:my_table5}
\begin{tabular}{| l  |l  |l  |l  |}
\hline
 & Al\textsubscript{2} (VQE) & Variation (±)& Al\textsubscript{2} (NumPy) \\
\hline
Circuit \#1 & -477.799 05& 0.000 000& -477.801 97\\
\hline
Circuit \#2 & -477.799 05& 0.000 000& -477.801 97\\
\hline
Circuit \#3 & -477.799 04& 0.000 000& -477.801 97\\
\hline
Circuit \#4 & -477.799 05& 0.000 000& -477.801 97\\
\hline
Circuit \#5 & -477.529 26& 0.097 844& -477.801 97\\
\hline
Circuit \#6 & -477.799 05& 0.000 000& -477.801 97\\
\hline

\end{tabular}

\end{table}

\begin{table}
\centering
\caption{Ground State Energy (\textit{E\textsubscript{h}}) Results of Varying Circuit Type on Statevector Simulator (Al\textsubscript{3}\textsuperscript{- })}
\label{tab:my_table6}
\begin{tabular}{| l  |l  |l  |l  |}
\hline
 & Al\textsubscript{3}\textsuperscript{- }(VQE) & Variation (±)& Al\textsubscript{3}\textsuperscript{- }(NumPy) \\
\hline
Circuit \#1 & -716.690 56& 0.000 000& -716.708 73\\
\hline
Circuit \#2 & -716.690 54& 0.000 000& -716.708 73\\
\hline
Circuit \#3 & -716.708 73& 0.000 000& -716.708 73\\
\hline
Circuit \#4 & -716.708 73& 0.000 000& -716.708 73\\
\hline
Circuit \#5 & -716.678 96& 0.019 614& -716.708 73\\
\hline
Circuit \#6 & -716.675 25& 0.000 000& -716.708 73\\
\hline

\end{tabular}

\end{table}

\begin{table}
\centering
\caption{Ground State Energy (\textit{E\textsubscript{h}}) Results of Varying Number of Repetitions (Reps) on Statevector Simulator (Al\textsuperscript{- })}
\label{tab:my_table7}
\begin{tabular}{| l  |l  |l  |l  |}
\hline
 & Al\textsuperscript{- }(VQE)& Variation (±)& Al\textsuperscript{- }(NumPy)\\
\hline
1 Rep & -238.842 31& 0.000 000& -238.870 47\\
\hline
2 Reps & -238.842 48& 0.000 000& -238.870 47\\
\hline
3 Reps & -238.857 88& 0.000 001& -238.870 47\\
\hline
4 Reps & -238.843 71& 0.000 007& -238.870 47\\
\hline

\end{tabular}

\end{table}

\begin{table}
\centering
\caption{Ground State Energy (\textit{E\textsubscript{h}}) Results of Varying Number of Repetitions (Reps) on Statevector Simulator (Al\textsubscript{2})}
\label{tab:my_table8}
\begin{tabular}{| l  |l  |l  |l  |}
\hline
 & Al\textsubscript{2} (VQE) & Variation (±)& Al\textsubscript{2} (NumPy) \\
\hline
1 Rep & -477.799 05& 0.000 000& -477.801 97\\
\hline
2 Reps & -477.799 04& 0.000 000& -477.801 97\\
\hline
3 Reps & -477.799 04& 0.000 002& -477.801 97\\
\hline
4 Reps & -477.799 04& 0.000 004& -477.801 97\\
\hline

\end{tabular}

\end{table}

\begin{table}
\centering
\caption{Ground State Energy (\textit{E\textsubscript{h}}) Results of Varying Number of Reps on Statevector Simulator (Al\textsubscript{3}\textsuperscript{- })}
\label{tab:my_table9}
\begin{tabular}{| l  |l  |l  |l  |}
\hline
 & Al\textsubscript{3}\textsuperscript{- }(VQE) & Variation (±)& Al\textsubscript{3}\textsuperscript{- }(NumPy) \\
\hline
1 Rep & -716.675 25& 0.000 000& -716.708 73\\
\hline
2 Reps & -716.708 72& 0.000 000& -716.708 73\\
\hline
3 Reps & -716.708 73& 0.000 000& -716.708 73\\
\hline
4 Reps & -716.708 72& 0.000 000& -716.708 73\\
\hline

\end{tabular}

\end{table}

\begin{table}
\centering
\caption{Comparison of Ground State Energy (\textit{E\textsubscript{h}}) Results of all Studied Systems with Varied Quantum Simulator Type (Statevector vs. QASM Simulator)}
\label{tab:my_table10}
\begin{tabular}{| l  |l  |l  |l|l|l|l|}
\hline
 & Al\textsuperscript{- }(VQE)&Al\textsuperscript{- }(NumPy)& Al\textsubscript{2} (VQE) & Al\textsubscript{2} (NumPy)& Al\textsubscript{3}\textsuperscript{- }(VQE)&Al\textsubscript{3}\textsuperscript{- }(NumPy) \\
\hline
Statevector & -238.842 31&-238.870 47& -477.799 05& -477.801 97& -716.675 25&-716.708 73\\
\hline
QASM & -238.310 78&-238.870 47& -477.206 89& -477.801 97& -716.470 32&-716.708 73\\
\hline

\end{tabular}

\end{table}

\begin{table}
\centering
\caption{Ground State Energy (E\textsubscript{h}) Results of Varying Basis Set on Statevector Simulator (Al\textsuperscript{-})}
\label{tab:my_table13}
\begin{tabular}{| l | l | l | l | l |}
\hline
 & Al\textsuperscript{- }(VQE)& Variation (±)& Al\textsuperscript{- }(NumPy) & Variation (±)\\
\hline
STO-3G & -238.842 31& 0.000 000& -238.870 47& 0.000 000\\
\hline
3-21G & -240.540 13& 0.000 406& -240.547 58& 0.000 655\\
\hline
STO-6G & -240.800 47& 0.000 020& -240.826 96& 0.000 479\\
\hline
6-31G & -241.844 63& 0.000 803& -241.851 93& 0.000 480\\
\hline
cc-pVDZ & -241.860 43& 0.000 235& -241.867 21& 0.000 908\\
\hline
cc-pVTZ & -241.860 32& 0.000 039& -241.866 67& 0.000 072\\
\hline

\end{tabular}

\end{table}

\begin{table}
\centering
\caption{Ground State Energy (E\textsubscript{h}) Results of Varying Basis Set on Statevector Simulator (Al\textsubscript{2})}
\label{tab:my_table14}
\begin{tabular}{| l | l | l | l | l |}
\hline
 & Al\textsubscript{2} (VQE)& Variation (±)& Al\textsubscript{2} (NumPy)& Variation (±)\\
\hline
STO-3G & -477.799 05& 0.000 000& -477.801 97& 0.000 000\\
\hline
3-21G & -481.076 34& 0.012 279& -481.095 08& 0.013 676\\
\hline
STO-6G & -481.710 46& 0.000 000& -481.713 39& 0.000 000\\
\hline
6-31G & -483.703 54& 0.000 732& -483.706 66& 0.010 760\\
\hline
cc-pVDZ & -483.737 01& 0.011 890& -483.741 61& 0.010 836\\
\hline

\end{tabular}

\end{table}

\begin{table}
\centering
\caption{Ground State Energy (E\textsubscript{h}) Results of Varying Basis Set on Statevector Simulator  (Al\textsubscript{3}\textsuperscript{-})}
\label{tab:my_table15}
\begin{tabular}{| l | l | l | l | l |}
\hline
 & Al\textsubscript{3}\textsuperscript{- }(VQE)& Variation (±)& Al\textsubscript{3}\textsuperscript{- }(NumPy)& Variation (±)\\
\hline
STO-3G & -716.675 25& 0.000 000& -716.708 73& 0.000 000\\
\hline
3-21G & -721.694 50& 0.000 000& -721.664 45& 0.377 995\\
\hline
STO-6G & -722.540 74& 0.000 000& -722.575 00& 0.000 000\\
\hline
6-31G & -725.601 86& 0.000 000& -725.603 34& 0.000 000\\
\hline

\end{tabular}

\end{table}